\documentstyle{mn} 
\input epsf.sty
\newif\ifAMStwofonts
\AMStwofontstrue
 
\ifoldfss
  \ifCUPmtlplainloaded \else
    \NewTextAlphabet{textbfit} {cmbxti10} {}
    \NewTextAlphabet{textbfss} {cmssbx10} {}
    \NewMathAlphabet{mathbfit} {cmbxti10} {} % for math mode
    \NewMathAlphabet{mathbfss} {cmssbx10} {} %  "   "    "
  \fi
  \ifAMStwofonts
    \ifCUPmtlplainloaded \else
      \NewSymbolFont{upmath} {eurm10}
      \NewSymbolFont{AMSa} {msam10}
      \NewMathSymbol{\upi}     {0}{upmath}{19}
      \NewMathSymbol{\umu}     {0}{upmath}{16}
      \NewMathSymbol{\upartial}{0}{upmath}{40}
      \NewMathSymbol{\leqslant}{3}{AMSa}{36}
      \NewMathSymbol{\geqslant}{3}{AMSa}{3E}

       \let\le=\leqslant
       
    \fi
  \fi
\fi % End of OFSS
 
\ifnfssone
  \newmathalphabet{\mathit}
  \addtoversion{normal}{\mahit}{cmr}{m}{it}
  \addtoversion{bold}{\mathit}{cmr}{bx}{it}
  \newmathalphabet{\mathbfit} % math mode version of \textbfit{..}
  \addtoversion{normal}{\mathbfit}{cmr}{bx}{it} 
  \addtoversion{bold}{\mathbfit}{cmr}{bx}{it}
  \newmathalphabet{\mathbfss} % math mode version of \textbfss{..}
  \addtoversion{normal}{\mathbfss}{cmss}{bx}{n}
  \addtoversion{bold}{\mathbfss}{cmss}{bx}{n}
  \ifAMStwofonts
    \ifCUPmtlplainloaded \else
and
your
      \UseAMStwoboldmath
      \makeatletter
      \new@mathgroup\upmath@group
      \define@mathgroup\mv@normal\upmath@group{eur}{m}{n}
      \define@mathgroup\mv@bold\upmath@group{eur}{b}{n}
      \edef\UPM{\hexnumber\upmath@group}
      \new@mathgroup\amsa@group
      \define@mathgroup\mv@normal\amsa@group{msa}{m}{n}
      \define@mathgroup\mv@bold\amsa@group{msa}{m}{n}
      \edef\AMSa{\hexnumber\amsa@group}  
      \makeatother
      \mathchardef\upi="0\UPM19
      \mathchardef\umu="0\UPM16
      \mathchardef\upartial="0\UPM40
      \mathchardef\leqslant="3\AMSa36
      \mathchardef\geqslant="3\AMSa3E

       \let\le=\leqslant

    \fi
  \fi
\fi % End of NFSS release 1
   
\ifnfsstwo
  \DeclareMathAlphabet{\mathbfit}{OT1}{cmr}{bx}{it}
  \SetMathAlphabet\mathbfit{bold}{OT1}{cmr}{bx}{it}
  \DeclareMathAlphabet{\mathbfss}{OT1}{cmss}{bx}{n}
  \SetMathAlphabet\mathbfss{bold}{OT1}{cmss}{bx}{n}
  \ifAMStwofonts
    \ifCUPmtlplainloaded \else
      \DeclareSymbolFont{UPM}{U}{eur}{m}{n}
      \SetSymbolFont{UPM}{bold}{U}{eur}{b}{n}
      \DeclareSymbolFont{AMSa}{U}{msa}{m}{n}
      \DeclareMathSymbol{\upi}{0}{UPM}{"19}
      \DeclareMathSymbol{\umu}{0}{UPM}{"16}
      \DeclareMathSymbol{\upartial}{0}{UPM}{"40}
      \DeclareMathSymbol{\leqslant}{3}{AMSa}{"36}
      \DeclareMathSymbol{\geqslant}{3}{AMSa}{"3E}

       \let\le=\leqslant

    \fi
  \fi  
\fi % End of NFSS release 2
      
\ifCUPmtlplainloaded \else
  \ifAMStwofonts \else % If no AMS fonts
    \def\upi{\pi}
    \def\umu{\mu}
    \def\upartial{\partial}
  \fi
\fi

\title{The structure and radiation spectra of illuminated accretion discs 
in
AGN. I. Moderate illumination}
\author[A. R\' o\. za\' nska, A.-M. Dumont, B. Czerny, S. Collin]
       {A. R\' o\. za\' nska$^1$, A.-M. Dumont$^2$, B. Czerny$^{1}$, S. 
Collin$^2$\\
        $^1$N. Copernicus Astronomical Centre, Bartycka 18, 00-716 Warsaw, 
        Poland\\$^2$ Observatoire de Paris, Section de Meudon, Place 
Janssen, 
92195 Meudon, France}
\begin{document}

\maketitle

\begin{abstract}

We present detailed computations of the vertical structure of an accretion 
disc 
illuminated by hard X-ray radiation with the code {\sc titan-noar} 
suitable for 
Compton thick
media. The energy generated via accretion is dissipated partially in 
the cold disc as well as in the X-ray source. 
We study the differences between the case where the X-ray source is in the
form of a lamp post above the accretion disc and the case of a heavy corona. 
We consider radiative heating via Comptonization together with heating
via photo-absorption on numerous heavy elements as carbon, oxygen, silicon,
iron. The transfer in lines is precisely calculated.
 A better description of the heating/cooling through the inclusion of 
line transfer, a correct description of the temperature in the deeper layers, 
a correct description of the entire disc vertical structure, as well as the
study of the possible coronal pressure effect,
constitute an
 improvement in 
comparison to previous works.
% including Nayakshin et al. and Ballantyne at al.

We show that exact calculations of hydrostatic equilibrium 
and determination of the disc thickness
has a crucial impact on the optical depth of the hot illuminated zone.
We study the lamp post model for a low ($\dot m = 0.03$) 
and high ($\dot m = 0.3$) accretion rate. In both cases we assume a 
moderate illumination where the viscous flux equals the X-ray radiation flux.
A highly ionized skin is created in the lamp post model, with 
the outgoing spectrum containing many emission lines and ionization
edges in emission or absorption in the soft X-ray domain, as well as an iron
line at $\sim 7 $ keV consisting of a blend of low ionization line from 
the deepest layers and hydrogen and helium like resonance line from the 
upper layers, and almost no absorption edge, contrary to the case of a slab of
constant density.
A full heavy corona completely suppresses the highly ionized zone on the top
of the accretion disc and in such case the spectrum is featureless. 

%Our model can be used for different accretion rates and 
%distributions of X-ray spectra, and can be compared with
%recent XMM and CHANDRA X-ray data. 

\end{abstract}


\section{Introduction}

Broad band spectra of active galactic nuclei (AGN) demonstrate the 
presence of both 
cold and hot plasma in the vicinity of a black hole. Furthermore, 
detailed analysis of X-ray spectra reveal that those  two phases interact 
radiatively
during their accretion on a central object 
(for  recent  review, see Collin et al. 2000, Reynolds 2000; 
see also Poutanen 1999 for the discussion of this 
problem in case of galactic black holes). 
However, the accretion pattern
is still a subject of dispute and several detailed scenarios were proposed.

In most cases, a key element of the model is an accretion disc which 
extends, or does
not, down to the marginally stable orbit. The disc surface is expected to 
be 
strongly irradiated by the X-ray emission containing a significant 
fraction 
of the observed bolometric luminosity of AGN. 

At a given disc radius, two distinct situations are possible. The
hot plasma may provide only a source of irradiating photons, being located 
well
above the disc or at different radius (the 'lamp post' model), 
or it may be located at
the considered radius, in hydrostatic equilibrium with the cold disc, 
additionally exerting a dynamical pressure on the cold disc surface 
(the 'heavy corona' model) and modifying strongly its ionization state 
(R\' o\. za\' nska et al. 1999).

In both families of models the optically thick disc is illuminated
by a hard X-ray flux. Those two cases, however,
differ significantly with respect to the ionization state of the disc 
surface
layers, as indicated by the study of R\' o\. za\' nska et al. (1999).
The density at the surface of a disc, irradiated through some non-locally
produced X-ray flux,
 is zero while the density of the disc with a corona is
quite high due to the coronal pressure. This may lead to significant 
differences
in the disc radiation spectra, both in the continuum and the properties of
the iron K$_{\alpha}$ line.

Careful approach to
the modeling of the broad band disc spectra is essential in order to test
both  models, particularly in view of the first chance for true X-ray
spectroscopy, coming with XMM and Chandra observations. 
Therefore, in this paper, we address the problem of the vertical structure 
and
the spectrum of the radiation  emitted  by the illuminated disc in two 
extreme 
cases: with
and without the dynamical effect of the corona.

For that purpose we consider the radiation transfer within the disc 
much more carefully than it was done by  R\' o\. za\' nska et al. (1999), 
applying
to the surface layers the code {\sc{titan}} developed by 
Dumont, Abrassart \& Collin (2000 hereafter DAC00).
This code also provides us with the continuum and line spectrum of the 
disc and
is interfaced with a Monte Carlo code {\sc{noar}} to take into account 
Compton diffusion
properly, and particularly to compute the iron K$_\alpha$ line profile and 
the high energy continuum.

This approach itself is an improvement in comparison to the earlier results
on the radiation spectra of X-ray irradiated accretion discs, although it 
is
still not free from important simplifications. 

We first stress that all computations of disc structure and disc spectrum 
are
(and will
   necessarily be)
based on simplifying assumptions because the problem is too complicated to
be solved exactly.
   And as it is not possible to make a 'uniform' improvement -
   it is worth to improve one aspect of the model at the expense of
another one. But different approximations leading to
different  predictions, it is necessary to perform a detailed comparison
  of the results, and this is not possible unless exactly the same model 
is used
in the
computation.
For instance Ballantyne et al.
(2001) and
Nayakshin et al. (2000) have computed the spectrum of
  an irradiated disc
  in hydrostatic equilibrium using codes similar to ours.
In the most simple case (an irradiated slab with a constant density), with
  exactly the same
parameters, the reflected spectrum predicted by the three codes
  shows important differences in the detailed spectral features as well as
  in the overall shape of the continuum
  (cf. Pequignot et al. 2001).
Thus in order
to judge the reliability of the results and to compare them to the
observations it is absolutely necessary that all authors be very clear on 
the
approximations made in their computations, so that everyone could be aware 
of
them.
   This is what we try to do in the present paper.

As improvements with respect to other methods, we have chosen here mainly
to favor the
radiative transfer treatment (other codes
  use the escape probability approximation, especially for the
lines), and to include a correct (but obviously model-dependent)
description of the vertical structure of the viscous disc underlying  the
irradiated layers (other codes use
  rough estimates from vertically averaged models). These improvments
are made at the expense of the atomic data. However though our
computations include less atomic transitions than
  XSTAR (used by Nayakshin et al. 2000), they include
more ions and transitions than Ballantyne et al. 2001.

In our approach the viscosity in the disc is proportional to the 
total (gas + radiation) pressure.
Close to the surface we compute the radiative transfer of continuum and 
lines 
using two-stream approximation with non-LTE treatment and with heavy 
elements
abundances as described in DAC00.  
Deeper inside the disc the diffusive radiative transport is assumed.
Due to better heating/cooling description we are able to 
calculate regions with temperatures below $\sim 10^5$, well matching with
the disc interior, i.e. full profiles 
of the disc from the surface to the equatorial plane.

%The presence of the physical corona, instead of just a source of hard 
%X-ray flux,
%is felt by the disc due to the dynamical pressure exerted by this hot 
%layer on
%the disc surface. It implies the change of the boundary conditions of 
%integrated 
%equations and affects the temperature and density profile.

We compute the emergent spectrum from the UV to the hard X-ray range, 
including 
features like the iron K$_{\alpha}$ line. 

This paper is restricted to a moderate illumination (the lamp post and
the heavy corona models), with low and high accretion rate, while  the 
following  paper will be devoted to the 
high illumination case (the flares).

In section 2 we present the assumptions and numerical method of 
computation. The results are presented in section 3.
The final sections are devoted to discussion and conclusions.

\section{Method}
\label{sec:met}

\subsection{Generalities}

We consider the vertical structure of an illuminated accretion disc
and the emitted spectra in two cases. In the first case hard X-ray emission
is produced in an active region located above an accretion disc 
(the lamp post model),
 and in the second case X-rays come from a hot corona (the heavy corona 
model).
The spectral distribution of X-rays is a free parameter of our model and 
in both cases it is assumed to be the same.
The only difference is that corona is heavy and exerts
a considerable pressure on
the disc surface thus modifying its properties.

In the case of lamp post model the surface
boundary conditions are the same as for a standard accretion disc, i.e.
the density (and gas pressure) approaches zero at the surface.

In the case of corona above an accretion disc the electron temperature 
of the hot plasma is chosen arbitrarily, in agreement with 
observations. We do not consider any particular coronal model in this 
paper.
The optical depth  and the gas pressure of the corona  are determined
according to the spectral distribution and the electron temperature 
from the relation that the flux emitted by the corona $F_{X}$ is produced 
via Compton cooling of hot electrons by the soft flux $F_{soft}$ outcoming 
from the accretion disc. The non-zero value of the coronal gas pressure 
implies 
surface condition on the disc atmosphere.

The disc structure in both cases is calculated in the same way, and the
presence or absence of the heavy corona influences the disc only through 
the
surface boundary conditions. These computations require an assumption  
about
the viscous energy generation within a disc. We adopt a standard $\alpha$ 
viscosity model where the viscous stress is proportional to the total 
pressure.

The results depend on the mass of the central black hole, $M$,
the distance from the 
center, $r$, the total flux generated by accretion, $F_{gen}$ (or total 
accretion 
rate $\dot m$), the fraction
of energy generated in the X-ray source, $f=F_{X}/F_{gen}$, and its 
spectral
shape, and finally on the viscosity parameter $\alpha$. 
Since  we do not specify any coronal model, we assume in both cases 
that all the angular momentum is transported outwards
by the disc, but only a fraction of the gravitational energy,
$ 1-f$, is dissipated in the disc. 
Therefore, the disc accretion rate is smaller than the total accretion 
rate which is
the parameter of our model, because  the flux dissipated in the disc, 
$F_{disc}$, is
smaller than the total generated flux; $F_{disc}=F_{gen}-F_{X}$. 
Note that in the case of lamp post model $f$ is not equal to $L_X/L_{bol}$.

The computations of the disc structure are done through iterations
between the code solving the hydrostatic equilibrium equation and the code 
solving the radiation transfer, as described below.

\subsection{The computation scheme}

\subsubsection{The beginning step}
\label{sec_firststep}

The first step of the computations is completed as described by R\' o\. 
za\' nska
et al. (1999 Section 2). 
The radiative transfer within the
disc is treated in the diffusion
approximation and the convective transport is also included. 
The viscous energy dissipation is given by the local $\alpha$ 
viscosity description. 
The hydrostatic equilibrium completes the set of equations
which fully determines the vertical temperature and the density profiles.

Thus, in the first step, we integrate the set of equations:
\begin{equation}
F=F_{\rm rad}= -{16 \sigma T^3  \over 3\kappa \rho }{dT \over dz},
\ \ \ \nabla_{\rm rad}\le\nabla_{\rm ad}, 
\end{equation}
\begin{equation}
F = F_{\rm rad} + F_{\rm conv}, \ \ \ \ \ \nabla_{\rm rad}>\nabla_{\rm ad},
\end{equation}
\begin{equation}
{dF \over dz} = {3\over 2} \alpha P \Omega_K (1-f) +{1\over 2}F_X \kappa 
\rho 
\exp(-\tau),
\label{eq:gen} 
\end{equation}
\begin{equation}
P = P_{gas} + P_{rad},
\end{equation}
\begin{equation}
{1 \over \rho} {dP \over dz} = - \Omega_K^2 z,  
\end{equation}
where: $F$, $F_{rad}$, $F_{conv}$, $T$, $\rho$, $z$, $P$, $P_{gas}$, 
$P_{rad}$,
$\Omega_K$, $\sigma$ and $\tau$ are respectively:
the flux emitted by the disc, the flux carried by radiation, the 
flux carried by 
convection,
the temperature, the density, the vertical coordinate, the total pressure, 
the gas pressure, 
the radiation pressure, 
the Keplerian angular velocity, the Stefan-Boltzman constant and the 
Rosseland mean 
optical depth.

The opacity $\kappa$ (the Rosseland mean) as a function of density and 
temperature is taken from Alexander, Johnson \& Rypma (1994) 
for log $T <3.8$, from Seaton et al. (1994) for  log $T >4.0$, and it is 
interpolated between these two tables for intermediate values of the
temperature.

For convection we adopt a simple description based on the mixing length theory 
used in
stellar interiors.

For both models we assume that half of the flux $F_{X}$ is directed toward
the disc and this fraction is  absorbed (the albedo equals zero in the first
computational step). Therefore, 
in both
cases we have the same boundary conditions on the flux and temperature at 
the 
surface. Using the  Eddington approximation we require:
\begin{equation}  
F(H_d) = F_{soft} \equiv 0.5 F_{X}+ (1-f) F_{gen},
\label{eq:sur1}  
\end{equation}
\begin{equation}
\sigma T^4(H_d) = {1\over 2}\sigma T_{eff}^4 ={1\over 2} F_{soft},
\end{equation}
where $H_d$ is half of the disc thickness actually determined by the 
boundary
condition on the equatorial plane:
\begin{equation}
F(z=0) = 0.
\end{equation}

The third boundary condition is different for both  models.
In the lamp post model, the density at the disc surface is:
\begin{equation}
\rho(H_d) =0.
\end{equation}

In the heavy corona model this condition results from the value of the
pressure at the basis of the
corona and the requirement of pressure equilibrium between the disc 
and 
the corona: 
\begin{equation}
\rho(H_d) =P_{gascor} {\mu m_H \over k_B T(H_d)}.
\end{equation} 
Assuming the vertical component of the gravity to be vertically constant,
the hydrostatic equilibrium  of the corona leads to  the gas pressure of 
corona:
\begin{equation}
P_{gascor}=\Omega_{K}^2 \frac{\tau_c}{\kappa_{es}} H_{cor}, 
\label{eq:gcor}
\end{equation}
where the pressure scale-height of the isothermal corona is of the order of:
\begin{equation} 
H_{cor}=\left( \frac{ k_B T_{cor}}{\mu m_H \Omega_{K}^2 }\right)^{1/2},
\label{eq:schei}
\end{equation}
where $\kappa_{es}$ is the opacity for electron scattering, $k_B$ - 
the Boltzman 
constant and $m_H$ - the mass of hydrogen. We assume the value of the mean 
molecular weight 
$\mu=0.5$ appropriate for cosmic chemical composition.  
Note that the pressure scale height $H_{cor}$ would be lower than in the 
case of 
variable gravity by a factor of $\sqrt{2}$.
Those two previous equations are equivalent, but they show 
clearly the dependence of $P_{gascor}$ and $H_{cor}$ on the free parameter 
$T_{cor}$.
The coronal plasma is taken to be a single temperature medium with 
$T_{cor}$ equal to the electron temperature chosen arbitrarily at 
$1 \times 10^9$ K (as suggested by observations of high energy cut off of 
X-ray 
spectrum). 
The optical thickness of the corona $\tau_c$ is not a free parameter of 
our 
model, but it is determined by the energy equilibrium. 
Since the 
corona is cooled via seeds photons coming from the 
accretion disc, it should have a  specific thickness to produce an X-ray
power-law spectrum with the energy index chosen in our computations.
We consider the case of an energy index $\alpha_E=0.9$, and we 
compute the coronal optical depth using a simple Comptonization 
code  (Czerny \& Zbyszewska 1991) based on
semi-analytical formulae 
and appropriate for optically thin media.  
 
We take into account only gas pressure due to the corona in this first 
step. 
The radiation pressure due to illumination is
included in the second step of iteration after solving radiative transfer.

We solve this two point boundary problem by a shooting method. 
The integration of equations presented in this section
is performed by the second order Runge--Kutta scheme with adaptive 
stepsize,
as in Pojma\' nski (1986).
 
\subsubsection{Radiative transfer in the surface layers}
\label{sec_rad}

Knowing the vertical density profile we can solve the radiative transfer 
within
the surface disc layers much more accurately using the code  
{\sc{titan/noar}} 
of 
DAC00. {\sc titan} is based on 
Eddington two-stream approximation of the
radiative transfer and works in plane-parallel geometry. 
Radiative transfer is computed both in lines and in continuum 
(i.e. lines are not treated with escape probability approach) so the
code can be used for very inhomogeneous thick media, as in the present 
case. 

Thermal equilibrium and  ionization and statistical equilibrium of ions 
are computed in
complete non-LTE for 10 elements and all corresponding ions.
We consider following elemental abundances:
H: $1$, He: $0.085$, C: $3.3 \times 10^{-4}$, N: $9.1 \times 10^{-5}$,
O: $6.6 \times 10^{-4}$, Ne: $8.3 \times 10^{-5}$, Mg: $2.6 \times 
10^{-5}$,
Si: $3.3 \times 10^{-5}$, S: $1.6 \times 10^{-5}$, Fe: $3.2 \times 
10^{-5}$,
from Allen (1973).
Comptonization is taken into account both in the thermal equilibrium and 
in the
computation of the spectrum, through an iteration with the Monte Carlo code
of Compton scattering, {\sc noar}.  

The radiative transfer equations for each frequency $\nu$ can be written 
in 
the form:
\begin{equation}
{1 \over {\sqrt 3}} {dI_{\nu}^{+} \over dz} = -(k_{\nu} + \sigma_{es} )
I_{\nu}^+ +  {\sigma_{es} \over 2}I_{\nu}^{-} + \eta_{\nu}
\end{equation}
\begin{equation} 
-{1 \over {\sqrt 3}} {dI_{\nu}^- \over dz} = -(k_{\nu} + \sigma_{es} )
I_{\nu}^- +  {\sigma_{es} \over 2}I_{\nu}^+ + \eta_{\nu} 
\end{equation}
where $z$ is the distance measured from the disc surface, $k_{\nu}$ is
the absorption coefficient in ${\rm cm}^{-1}$, $\sigma_{es}$ is
the electron scattering coefficient, 
and $\eta_{\nu}$ is the emissivity.

The boundary conditions are imposed at the illuminated side of the disc 
surface and asymptotically in the disc interior.

The surface boundary condition is given by:
\begin{equation}
I^+_{\nu}(0)={{\sqrt 3} \over 2 \pi} F_{\nu}^{X}
\end{equation}
where $F_{\nu}^{X}$ is the frequency dependent illuminated X-ray flux.
 
It is not necessary to continue with integrations down to the 
equatorial plane of the disc since non-LTE effects disappear completely 
when the optical depth is $\sim 10$ , and where the diffusion
approximation is satisfactory anyway. 

Therefore we divide the disc arbitrarily into a surface
layer of thickness $d$ and the interior, like in the case of a stellar 
atmosphere problem. The disc interior is calculated in
the diffusion approximation of radiative transfer, and the zone
down to $z=d$ is computed by {\sc{titan}}. However, we have to impose
a second boundary condition at $z=d$ which will insure that the energy flux
dissipated in the disc $F_{\nu}^{disc}(d)$ will leave the interior:

%Boundary conditions are formulated
%on both surfaces. At the X-ray illuminated side we give the 
%total illuminating
%X-ray flux, $F_X$, and we specify its spectral shape or compute it from 
%the 
%corona model. In two-stream approximation outer boundary condition
%for illuminated atmosphere is 
\begin{equation}
I^{-}_{\nu}(d)=I^{+}_{\nu}(d)+ {{\sqrt 3}\over 2 \pi} F_{\nu}^{disc}(d).
\end{equation} 

Since the disc computations provide us only with the frequency-integrated 
value
of the flux $F_{disc}$ dissipated in the disc 
and the value of the temperature at $z=d$, we determine
the spectral shape of the flux using the non-grey diffusion approximation 
but
neglecting the frequency dependence of the opacity coefficient
\begin{equation}
F_{\nu}^{disc}(d)={4 \pi \over 3}{dB_{\nu}(T) \over d \tau} = F_{disc}(d)
{\pi B_{\nu}(T)\over 4 \sigma^4}{x \exp(x) \over( \exp(x) - 1)},
\label{flu:disc}
\end{equation}
where $x=h\nu/k_B T$, $B_{\nu}(T)$ is the Planck function.
This condition is equivalent to the condition used by NKK00 (Eq.14) in 
their
illuminated disc model.

The code returns the temperature profile and the spectral shape of the
emitted radiation.
The results are not sensitive to the choice of  $d$ as long as 
$\tau_{es}(d)$ is 
neither too small (smaller than $3$)  nor too large (larger than $30$)
since in the first case the
radiative transfer of the incident X-ray flux is not described
accurately, and in
the second case the accumulated error of the
computations made practically in
LTE zone is too large. We arbitrarily choose $\tau_{es}(d) \sim 6$. 
In earlier works it was estimated that X-rays penetrate the 
disc atmosphere up to an optical depth approximately equal  $3$ (Sincell 
\& Krolik
1997, NKK00)
 
\subsubsection{Density profile from hydrostatic equilibrium}
\label{sec_den}

Having the new temperature profile of the surface layers from the code 
{\sc{titan}}
calculated for the density profile in the first step, we can 
calculate a new density profile from hydrostatic equilibrium. For that
purpose we solve  the entire disc structure basically as described in 
Section~\ref{sec_firststep},
but with few modifications.
At a distance smaller than $d$ from the disc surface we assume the
temperature profile given by Section~\ref{sec_rad} and we do not solve the
diffusive 
radiative transfer (Eqs. 1 - 3), while deeper within the disc we solve the 
full disc vertical
structure equations including diffusive radiative transfer and heat 
generation. 
At this stage we also include the radiation pressure due to the 
incident radiation. Having the radiation field given by  {\sc{titan}}
we determine this pressure component by:
\begin{equation}
\frac{d P_{rad}}{dz}= \frac{1}{c {\sqrt3}} \int k_{\nu} F_{\nu} d\nu,
\end{equation}
where the radiation pressure on the surface is equal to $F_{X}/c$.

In the heavy corona model the value of the density at the disc
surface changes in subsequent iterations due to the change
of the surface temperature, since it is determined from the adopted value 
of
the pressure at the basis of the corona.
This differs from the case of the lamp post model where the surface 
density is 
zero through all iterations. 

One full iteration is done after computing the temperature and density 
profiles.
The new density profile is used to repeat radiative transfer computations
as described in Section~\ref{sec_rad}. We again divide the disc in a
surface layer with optical thickness $\tau_{es}(d) \sim 6$, and the
rest of the disc,
where the diffusion approximation is adopted. 

We repeat iterations between radiative transfer and density profile 
until the temperature profile does not change, and in section
below we present the results.    

\subsubsection{Iron line}

When the irradiated disc structure is determined we calculate in more 
detail
the shape of the X-ray spectrum in order to determine more accurately
the properties of the iron K$_{\alpha}$ line. These computations are done
by the Monte Carlo code {\sc noar} described in detail by DAC00.
This code describes the radiative transfer of line components coming from 
various iron ions, as determined by {\sc titan}, and includes the line
broadening due to Comptonization. It is used also after each {\sc titan}
run to provide the net (heating - cooling) Comptonization effect. 
We do not include here any kinematic
broadening related do the systematic motion of emitting material thus 
providing 
an 'intrinsic spectrum' which later may be folded with arbitrary flow 
pattern like Keplerian disc motion, inflow or outflow.

\section{Numerical results}

All results are presented for a single 
representative radius $r = 10 R_{Schw}$ (where $R_{Schw}=2GM/c^2$), 
for a mass of black hole 
$M=10^8 M_{\odot}$, two accretion rates (in units of Eddington accretion 
rate with
efficiency $1/12$): $\dot m = 0.03$ and $\dot m = 0.3$, a viscosity 
parameter $\alpha=0.1$, and
for $f=0.5$.  For these parameters, the total X-ray flux 
as well as locally dissipated flux are equal to 
$6.96 \times 10^{13}$ erg s$^{_1}$ cm$^{-2}$, with an incident X-ray 
radiation flux equal to half of this value.

The distribution of incident X-ray spectrum is a power-law with 
energy index $\alpha_{E}= 0.9$, extending 
from 2.8 eV up to 100 keV.

\subsection{Vertical structure of the disc without corona}
\label{sec:active}
We perform the coupled computation of the hydrostatic equilibrium and the 
radiative transfer for an irradiated accretion disc as 
described in Sect.~\ref{sec:met}.

\begin{figure}
 \epsfxsize = 80mm \epsfbox[25 300 550 700]{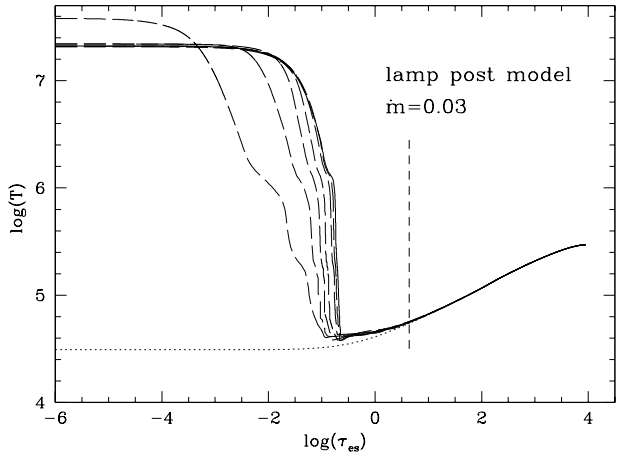}
 \caption{Temperature profile versus Thomson optical depth
for the lamp
post model. The initial profile is denoted by the dotted line, next 
iterations
are presented as long dashed lines. 
The final profile is shown by the solid line. The vertical straight line marks 
the
transition between the diffusion approximation and the exact solution.}
 \label{fig:iter}        
\end{figure}

We do not include conduction but the iterative method allows to pass
effectively  from the
hot upper layers to the cold inner disc without serious instability 
problems.
Contrary to the computations done by NKK00 we do not start calculation of 
radiative
transfer from an already unstable temperature 
profile determined with the  photoionization code.
%Therefore, {\bf thermaly unstable regions}, whenever they take place, 
%should 
%appear during computations of our radiative transfer.  
The successive iterations of the temperature profile for $\dot m = 0.03$ 
are shown in 
Fig.~\ref{fig:iter}
starting from the solution given by diffusion approximation (dotted line).

The temperature distribution displays an expected pattern; the hot upper 
layer reaches approximately the inverse Compton temperature
(determined by the spectrum of the incident radiation and the radiation 
spectrum emitted by the disk and its atmosphere), then follows a
rapid temperature drop due to an increase of the density and, 
consequently, an
increase of
the cooling efficiency of the gas. 
The deeper layers of the disc are not affected by the irradiation, so the 
better
description of the radiative transfer than in the paper of 
R\'o\.za\'nska et al. (1999)  does not alter their results for the
disc interior. After an initial decrease close to the surface, the 
temperature 
inside the disc is rising towards the equatorial plane, reaching 
$T \sim 2.94 \times 10^5 $ K.
The solution converges satisfactorily after
 seven iterations. The number of iterations cannot be easily 
increased as the 
computations are time-consuming, taking about 3 days per iteration. 

\begin{figure}
 \epsfxsize = 80mm \epsfbox[25 300 550 700]{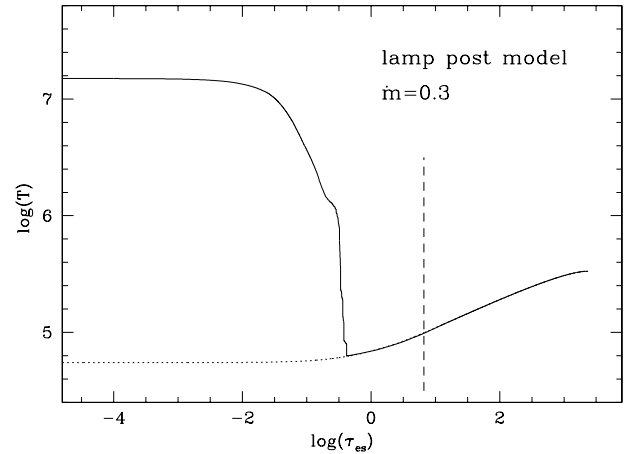}
 \caption{Temperature profile versus Thomson optical depth
 for the lamp
post model. The initial profile is denoted by the dotted line and the 
final profile is shown by solid line. The vertical straight line marks the
transition between the diffusion approximation and the exact solution.}
 \label{fig:iter2}        
\end{figure}

The temperature profile for the case of high accretion rate
is presented in Fig.~\ref{fig:iter2}. The  heated zone is thicker in 
this case than in the lower accretion rate solution. 
 
In the final stage of iteration with radiative transfer we obtain an 
optical depth of
the Compton heated skin $\tau_{es}^{hot} \sim 0.03$ 
(defined by a decrease of the temperature by a factor 2 with respect to 
the surface temperature, in reference to the simple approach of Begelman, 
McKee and Shields 1983, the discussion of McKee and Begelman 1990 
and Appendix A)  for $\dot m = 0.03$ 
and $\tau_{es}^{hot} \sim 0.048$ for $\dot m = 0.3$
(see Figs.~\ref{fig:iter} and \ref{fig:iter2}). 
For $\dot m=0.03$ the height of the bare disc is $H_d\sim 3.7 \times 
10^{12}$ cm, 
and the height of the hot skin is  $H_{hot}\sim 5.1 \times 10^{12}$ cm.
The transition layer is very thin, with $H_{tran}=6 \times 10^{11}$ cm.
The same geometrical sizes in case of $\dot m = 0.3$ are as follows: 
$H_d\sim 2.3 \times 10^{13}$, $H_{hot}\sim 1.44 \times 10^{12}$ , and 
$H_{tran}=1.1 \times 10^{11}$ cm.
These  values are much smaller than the disc radius ($3\times 10^{14}$ cm)
which means that both the disc and the hot layer are geometrically thin.

\begin{figure}
 \epsfxsize = 80mm \epsfbox[25 300 550 700]{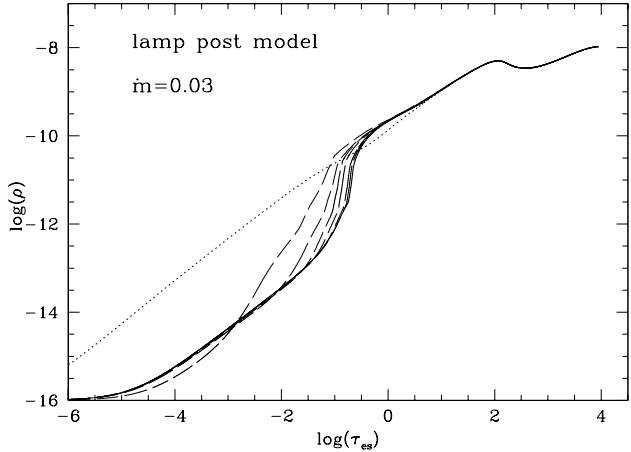}
 \caption{Density distribution versus Thomson optical depth for the lamp
post model. The initial 
profile is denoted by dotted line, next iterations are presented as long 
dashed 
lines. Final profile is shown by the solid line.}   
\label{fig:gest}     
\end{figure}

\begin{figure}
 \epsfxsize = 80mm \epsfbox[55 30 760 500]{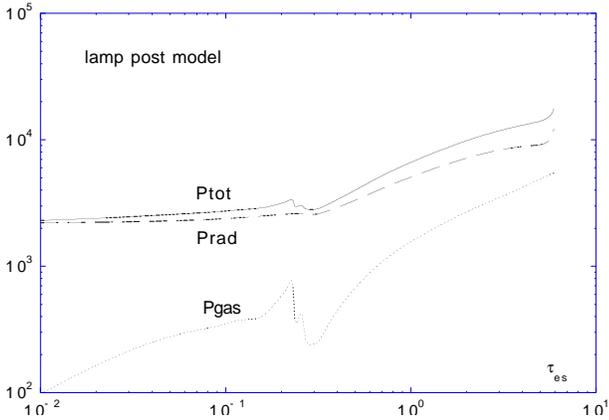}
%\epsfxsize = 85 mm
%\epsfbox{fig-agaw7-P.eps}
\caption{The pressure distribution
 versus Thomson optical depth in the surface layers 
for lamp post model after the last
iteration for $\dot m =0.03$.} 
\label{fig:cis}             
\end{figure}

Equation~\ref{eq:schei} gives analytically the scale-height of the 
Compton heated
zone, when $T_{cor}$ is replaced by the inverse Compton temperature 
$T_{IC}$.
For the accretion rate $\dot m = 0.03$ and $T_{IC}=2.1\times 10^7 $ K, as 
calculated from the spectral distribution of hard X-rays and
the disc  radiation, we can analytically derive the value
of $H_{hot} = 2.6 \times 10^{12}$ cm.
We see that the actual thickness of the Compton skin is a  
factor of $2$ larger when exact vertical calculations are done. 
The geometrical extension of the hot zone formally depends on the adopted 
value
of the initial small (but non-zero) density at the surface but in practice 
this
dependence is weak since the gravity in the disc increases outwards.

In Fig.~\ref{fig:gest} we show the density profile {for $\dot m = 0.03$}, 
again 
starting from the diffusion approximation.
The density of the illuminated layer is lower than for the bare disc
up to a Thomson optical depth $\sim 0.1$. 
Deeper in the disc, the density profile causes a density inversion
due to convection, as shown in R\'o\.za\'nska et al. (1999).

The final gas pressure and  
total pressure profiles for $\dot m=0.03$ 
are shown for surface zones in Fig.~\ref{fig:cis}.
Radiation pressure always dominates the gas pressure by at least two 
orders of
magnitude. 

The pressure profile resulting from diffusion approximation
goes monotonically through the disc.
Nevertheless in  outer layers we  see a 'wiggle' in the region of the 
temperature drop but the hydrostatic equilibrium and the radiative 
transfer are nevertheless computed
with the required accuracy at each step.
We argue here, that such  'wiggles' are due to thermally unstable regions,
which are expected to appear in an atmosphere in hydrostatic 
equilibrium and illuminated
by hard X-rays (RC96).

Since we start our computation from the thermally stable temperature 
profile
of the non illuminated disc atmosphere, and since we perform an iteration 
between this profile and a single value density profile, we do not 
achieve a three value temperature profile (i.e. thermally unstable regions)
after the radiative transfer computation.
Note that iterating between temperature and density profiles is a usual 
way of
treating any radiative transfer computations in hydrostatic equilibrium
(see Hubeny 1990, Madej \& R\'o\.za\'nska 2000, NKK00). 
In thermally unstable regions, pressure is a monotonic function
of optical depth (RC96), and therefore
if we keep temperature as a single value function of $\tau_{es}$ in our 
radiative transfer computations,
the pressure inversion will appear in the thermally unstable region. We 
note, however, that the wiggle occupies only a small part of the 
irradiated layer and in particular it does not affect the overall shape of 
the emitted spectrum but it may have some impact on the detailed spectral 
features (cf. Sect.~\ref{sect:instab}).   

As a conclusion we stress that in any radiative transfer computation of
disc atmospheres illuminated by hard X-rays, if the initial  temperature
profile does not displays instabilities, the final pressure profile 
will display inversions. 
It may be the way to check that the cooling and heating are correctly 
computed,
as the instabilities are due to discontinuities in the cooling rate 
through the 
illuminated atmosphere. Including the conduction flux into the scheme would
solve the problem but it is technically difficult and no exact radiative 
transfer
 computations were performed with the conduction effect included up to now.

\subsection{Outgoing spectrum in the lamp post model}

\subsubsection{Overall spectral distribution}

\begin{figure*}
 \epsfxsize = 152mm \epsfbox[40 310 470 670]{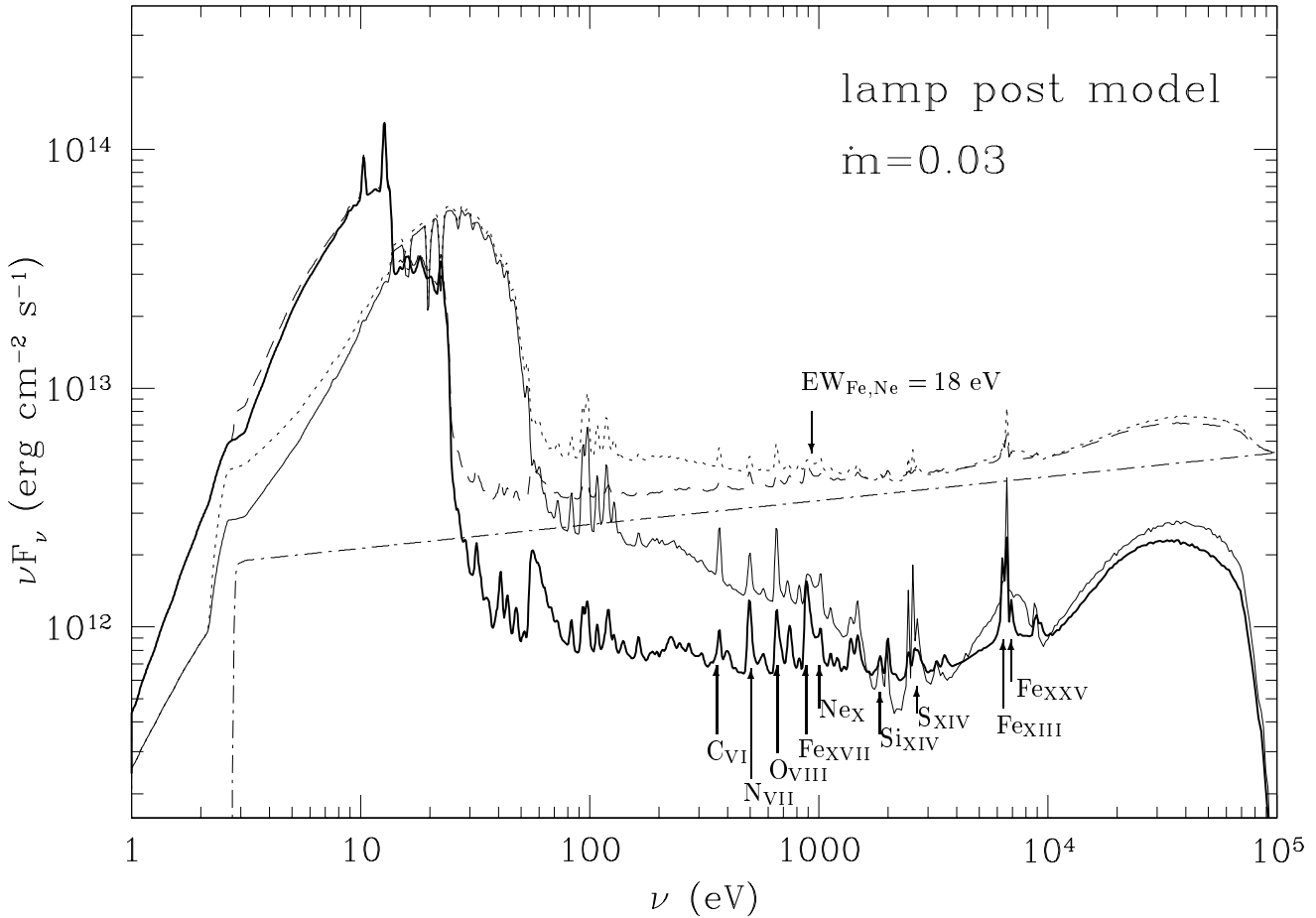}
 \caption{The reflected spectra of the irradiated disc for the lamp post 
model
for hydrostatic equilibrium (thick solid line), and for constant density 
model 
(thin solid line).
Final spectra are presented by dotted and dashed lines respectively.
Power law line marks spectral distribution of incident radiation. Spectral resolution: 30.}
 \label{fig:spec1}        
\end{figure*}

The reflected spectrum for $\dot m = 0.03$, obtained from the final iteration,
 is presented as a thick solid
line in 
Fig.~\ref{fig:spec1}, which gives also the incident 
illuminating spectrum (power law line).
The dashed line shows the observed spectrum, equal to the sum of
the incident and the reflected spectra. Plotted spectra are usually
degraded to a spectral resolution of $R = 30$, and no kinematic corrections are
applied.

The reflected spectrum consists of a large optical/UV/soft X-ray component 
due to the disc emission as well as to thermalization of a significant 
fraction 
of the incident radiation. The X-ray part of the spectrum shows the 
characteristic
spectral shape of reflection from partially ionized medium, with
much more radiation in the soft X-ray band than in the case of a neutral
medium (Lightman \& White 1988), and in particular a lot of emission lines 
and
ionization edges.

The reflected spectrum is very different from that of a constant 
density slab. It is actually not easy to compare precisely the two 
cases, as for hydrostatic equilibrium the ionization parameter has no 
meaning and it is also difficult to define an average density.  To 
perform the comparison, we have used a slab of constant density equal 
to  10$^{12}$ cm$^{-3}$ with {\it exactly} the same irradiation as in 
the hydrostatic case: a power law spectrum with the same flux 
incident on the top of the layer, and a thermal spectrum incident on 
the bottom of the layer, with $F_{\nu}^{disc}$ given by 
Eq.~\ref{flu:disc}. 
This is comparable to the method used  for instance by Ross et 
al (1993) and \.Zycki et al (1994) to compute the disc spectrum, except 
that they had a larger density, representative of the mid-plane 
value, while we have a density more representative of the irradiated 
layers. The resulting reflected spectrum is displayed as a thin solid
 line on Fig.~\ref{fig:spec1}, in order to be compared with the lamp 
post  reflected spectrum, and the corresponding observed spectrum is 
shown as a thin dotted line.

\begin{figure} 
 \epsfxsize = 75mm \epsfbox[50 150 480 780]{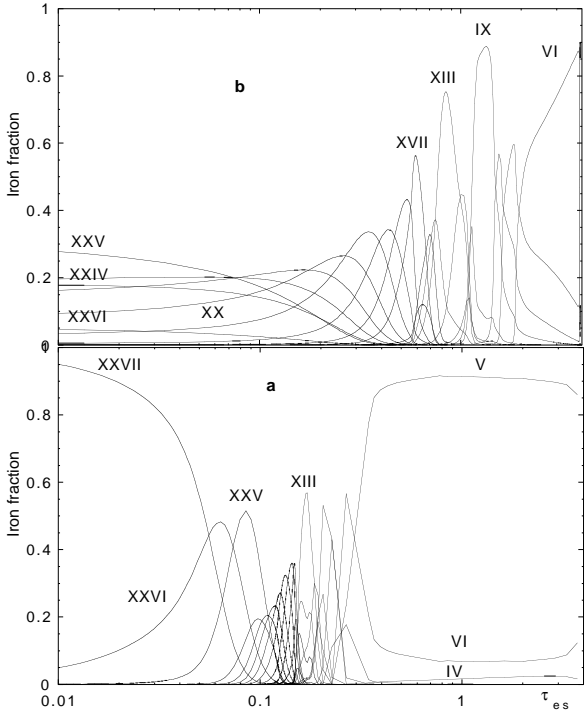}
%\epsfclipon \epsfbox{suz2.2.eps}
 \caption{The fraction of iron ions in case of lamp post model versus
Thomson optical depth in constant density case (upper panel) and 
hydrostatic equilibrium (lower panel). Accretion rate: $\dot m = 0.03.$}
 \label{fig:ion}        
\end{figure}

We see that the constant density case shows a larger thermal UV bump, 
very close to the underlying black body continuum, with only small 
traces of ionization edges in absorption, while the hydrostatic model 
exhibits a Lyman edge in absorption and a strong absorption above 20 
eV, caused by the presence at $0.1 \le \tau_{es}\le 1$ of several 
weakly ionized species: O {\sc ii} and O {\sc iii},  S {\sc ii} and 
S {\sc iii}, C {\sc iii}, Si {\sc ii}  and 
Si {\sc iii} ions.  As an illustration Figs.~\ref{fig:ion}  a and b  show 
the 
fractional abundances of the different iron ions as functions of the 
optical depth, respectively for the hydrostatic equilibrium and for 
the constant density case: in the latter case we see that only 
relatively highly ionized species are present in the deep layers, 
contrary to the hydrostatic case. In both cases highly ionized 
species are present at $\tau_{es}\le 0.1$.

The low ionization state in the hydrostatic case is due to the high 
density of the deep layers. Since LTE is almost reached, the 
ionization stage depends only on the density and the temperature 
profile. It is therefore {\it most important to solve the whole 
vertical structure of the disc to determine the abundances of these 
weakly ionized species and to get the correct spectral distribution 
in the far UV 
range}.

In the soft X-ray range the shape of the reflected spectra differs 
also in the two cases, as it is steeper for the constant density. 
This is due to the lack of thermal instability and consequently 
the smooth distribution of the temperature in the constant density case. 
On the contrary, the shape of the reflected spectrum in the hard X-ray 
range is almost the same in the two cases, as the Thomson thickness 
of the hot layer is relatively small ($< 0.1$) so Compton reflection 
takes place in a cold medium. Also because of the small optical 
thickness of the hot skin, the outgoing spectrum is not modified by 
Comptonization in the UV and soft X-ray ranges. Finally, one should 
notice that in the soft and hard X-ray range the overall shape of the 
{\it observed} continuum is quite similar in the two cases, and it is 
only the (very different) detailed spectral features which could help 
to distinguish between them. In particular, the spectrum in the 
hydrostatic case exhibits strong ionization edges in emission in the 
soft X-ray range, but {\it no ionization edge of Iron} in absorption 
at $\sim 10$ keV, contrary to the constant density case (see 
Figs.~\ref{fig:spec1}
and \ref{fig:lin}); one can see only a
weak Fe{\sc xxvi} ionization edge in emission.
Computations performed under assumption of constant pressure give
results intermediate between the constant density and hydrostatic 
equilibrium 
solutions (Dumont et al. 2001).

\begin{figure}
 \epsfxsize = 80mm \epsfbox[25 300 550 700]{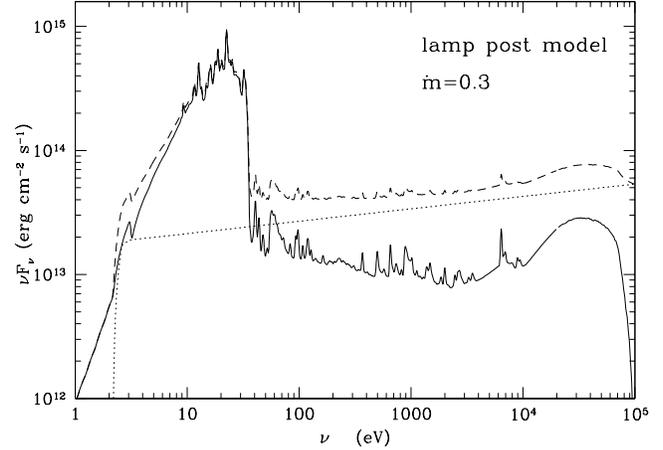}
 \caption{The reflected spectrum (continuous line) and final spectrum 
(dashed line) for the lamp post model in hydrostatic equilibrium for
high accretion rate. Spectral resolution: 30.}
 \label{fig:spechigh}        
\end{figure}

The effect of the adopted  value of the accretion rate can be seen 
comparing  Figs.~\ref{fig:spec1} and \ref{fig:spechigh}. The spectra are qualitatively 
similar, but they also show systematic differences matching the difference
in the temperature profile.

\begin{figure}
 \epsfxsize = 75mm \epsfbox[50 25 690 550]{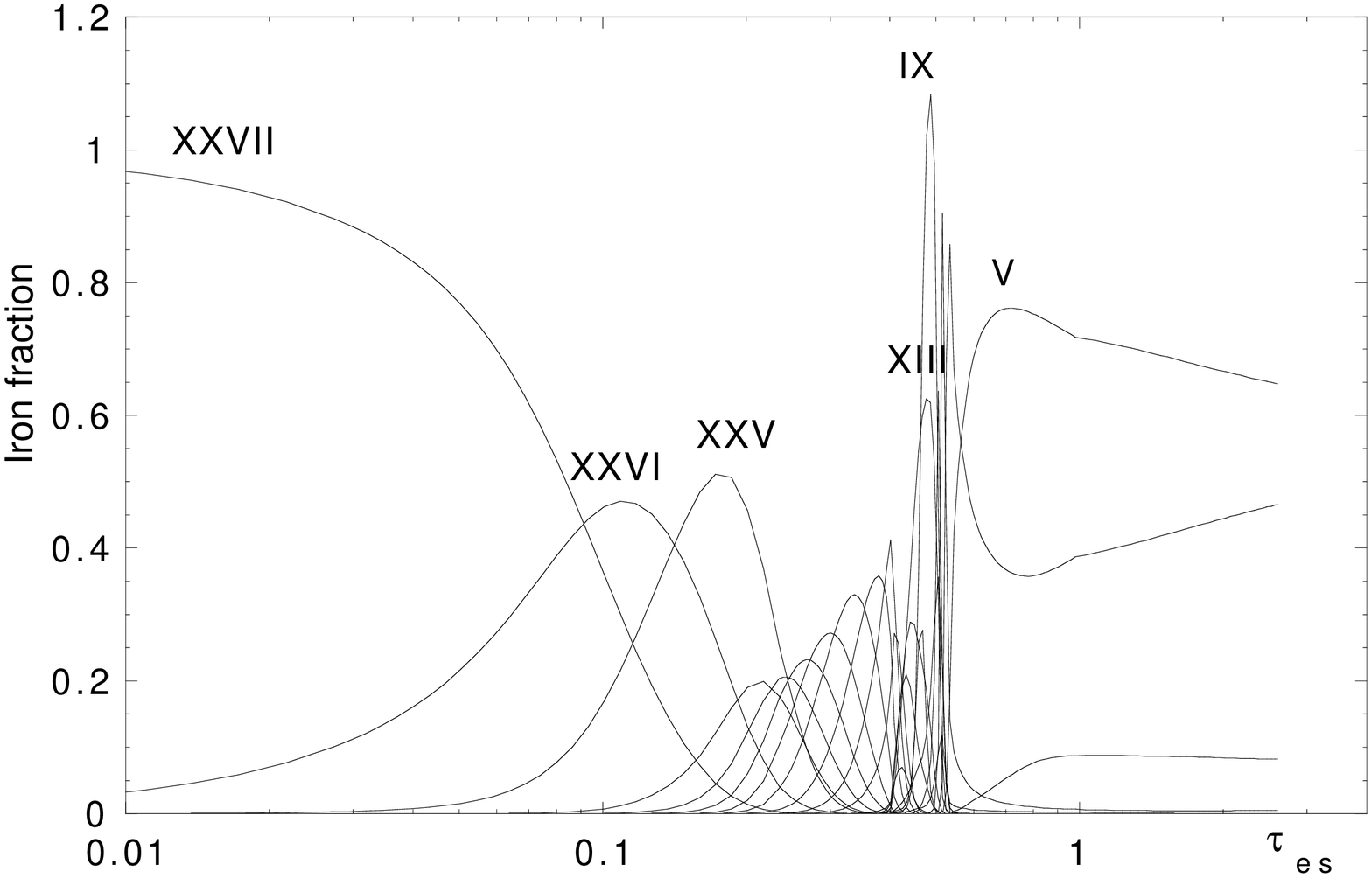}
 \caption{The fraction of iron ions in case of lamp post model versus
Thomson optical depth in 
hydrostatic equilibrium. Accretion rate: $\dot m = 0.3$.}
 \label{fig:ionhigh}        
\end{figure}

The overall reflectivity in the X-ray band is higher for higher accretion rate,
as it corresponds to somewhat higher ionization state of the gas. This is well
seen in the plot of fractional abundances of iron 
(see Fig.~\ref{fig:ionhigh}). FeXXV
is emitted predominantly from the optical depth of $\sim 0.09$ for 
$\dot m = 0.03$ 
but from the optical depth of $\sim 0.18$ for $\dot m = 0.3$ while  FeV comes 
from $\tau_{es} > 0.3$ and  $\tau_{es} > 0.7$, correspondingly.
There is 
systematically even more reflection in the soft X-ray band, making this
case  more 
similar to the constant density case than it was for lower accretion rate.
Also for $\dot m =0.3$ there are copious soft X-ray lines from CNO elements
and other species, although they are generally less intense than for 
$\dot m = 0.03$ if measured with respect to the reflected component 
but have similar
EW measured with respect to the total continuum. Higher average albedo leads
to the reduced dominance of the optical/UV bump.

\subsubsection{K$_{\alpha}$ line properties}

\begin{figure}
 \epsfxsize = 75mm \epsfbox[60 50 700 520]{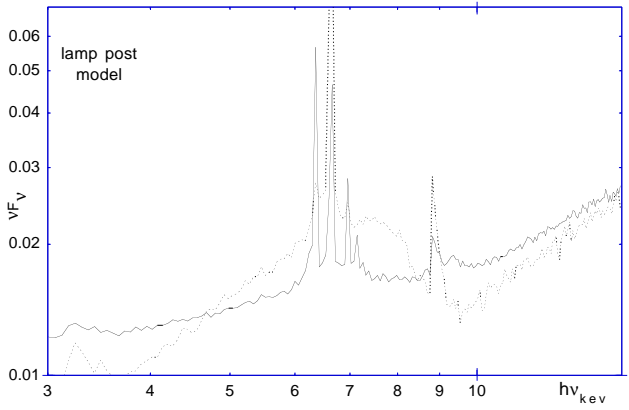}
 \caption{The close-up of the iron line region of the reflected spectrum 
(in arbitrary units)
for lamp post geometry and constant density case 
(dotted line) and hydrostatic equilibrium (solid line). Spectral resolution
$R = 100$. Accretion rate: 
$\dot m = 0.03.$}
 \label{fig:lin}        
\end{figure}

\begin{figure}
\epsfxsize = 75mm \epsfbox[75 200 480 510]{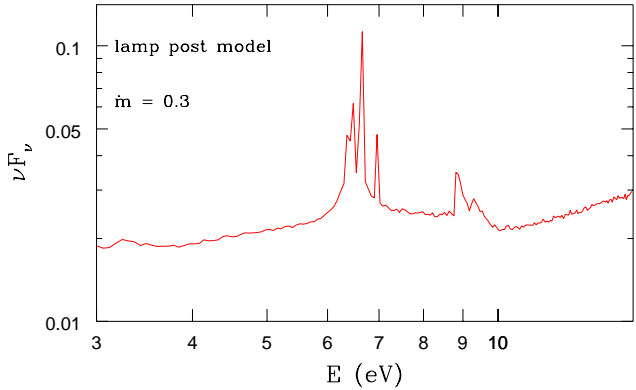}
 \caption{The close-up of the iron line region of the reflected spectrum 
(in arbitrary units)
for lamp post geometry and hydrostatic equilibrium (solid line). 
Spectral resolution
$R = 100$.}
 \label{fig:linhigh}        
\end{figure}

The differences of ionization state in the two cases induce 
quite different Iron  spectrum, as already stressed by NKK00 
(who have performed a comparison using a constant 
irradiation flux and not a constant $F_X/F_{visc}$ ratio like here). 
Figs.~\ref{fig:lin} 
and ~\ref{fig:linhigh} display  enlarged 
versions of the reflected spectra between 3 and 15 keV with relatively high 
spectral resolution. We see that in 
the hydrostatic case the spectrum is dominated by two components: 
the 6.4 keV component is due to several Iron K$\alpha$ lines of Fe{\sc xiii} 
and less ionized species (the most intense line being due to Fe{\sc v}), 
and the 6.7 keV component is the Fe{\sc xxv} recombination line blend 
with
Fe{\sc xxiv}. For $\dot m = 0.03$ the first one slightly dominates while
for $\dot m = 0.3$ the second one is stronger. The 
recombination line of Fe{\sc xxvi} is also present at 6.9 keV but is much 
weaker, as well as the K$_\beta$ line at 7.1 keV. An emission edge due 
to Fe{\sc xxv} is seen at 8.8 keV. 
It is difficult to compute the equivalent width EW of each
component separately, as they are blended together through Compton
broadening. The whole line complex has an EW of 550 eV with respect
to the reflected continuum, corresponding to an EW of 90 eV with respect
to the observed continuum in the case of the lower accretion rate.
For high accretion rate, the corresponding numbers are 405 eV and 90 eV. 
Dividing roughly the whole line profile
 into the three
lines, we obtain the values of the EW of the line components given with respect to reflected and total observed continuum, 
correspondingly (see Table~\ref{tab:EW}).

%:
%\begin{itemize}
%\item EW(cold Iron): 240 eV  and  40 eV
%\item EW(FeXXIV+FeXXV): 220 eV  and 37 eV
%\item EW(FeXXVI): 80 eV  and 15 eV.
%\end{itemize}

  \begin{table}
  \caption{Iron $K_{\alpha}$ line properties in the considered models$^*$.
  \label{tab:EW}}
  \begin{tabular}{l|rr|rr}
   \hline
  &  \multicolumn{2}{c}{$\dot m = 0.03$}&\multicolumn{2}{c}{$\dot m = 0.3$}\\
\hline  
component   &  $EW_{refl}$ & $ EW_{obs}$  & $ EW_{refl}$ & $ EW_{obs}$ \\
\hline
cold Iron &  240 eV & 40 eV & 145 eV & 32 eV \\
FeXXIV+FeXXV & 220 eV & 37 eV & 220 eV & 49 eV\\
FeXXVI & 80 eV & 15 eV & 40 eV & 9 eV \\  
\hline
   &       &      &    &                 \\
  \end{tabular}
$^*$  $F_x/F_d = 1$, $\Gamma_{PL} = 2$, $E_{max} = 100$ keV, $M = 10^8 M_{\odot}$, $ r = 10 R_{Schw}$
  \end{table}
 
These numbers are approximate: in particular the EW of the FeXXVI
line is overestimated, as it integrates a blue wing partly due to
inverse Compton broadening of the other lines, and as a consequence
the other lines are slightly underestimated.

In the 
constant density case the spectrum is dominated by the Fe{\sc xxv} line 
and 
by a corresponding intense ionization edge in emission. There is also 
a smooth but strong absorption above 8 keV due to the superposition 
of several ionization states.

\subsection{The weight of the corona effect}

\subsubsection{Vertical structure of the disc with coronal influence}

In the case of the heavy corona we assume the electron temperature of the 
hot plasma to 
be equal $1 \times 10^9$K as expected from observations. This temperature 
and 
the assumed spectral distribution of X-rays imply an optical depth of 
corona 
$\tau_{cor} = 0.355$ due to the role of Comptonization (Haardt \& Maraschi 
1991).
Therefore, the coronal pressure, calculated from Eq.~\ref{eq:gcor}, is 
equal to 
$8200$ in cgs units.
All other parameters are the same as described in Sect.~\ref{sec:active}.

The dynamical pressure of the corona modifies the disc atmosphere, 
making the gas relatively cool and dense (R\'o\.za\'nska et al. 1999). 

\begin{figure}
 \epsfxsize = 80mm \epsfbox[25 300 550 700]{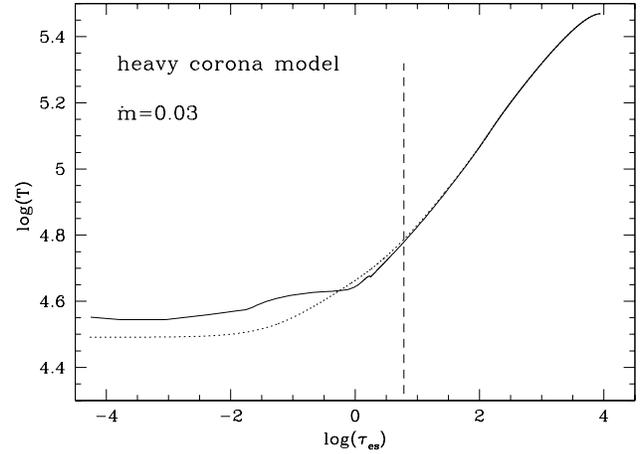}
 \caption{The temperature distribution for the heavy 
corona
          model. Initial profile is represented by dotted line, 
and final iteration by solid 
line.}
 \label{fig:iter3}        
\end{figure}

\begin{figure}
 \epsfxsize = 80mm \epsfbox[25 300 550 700]{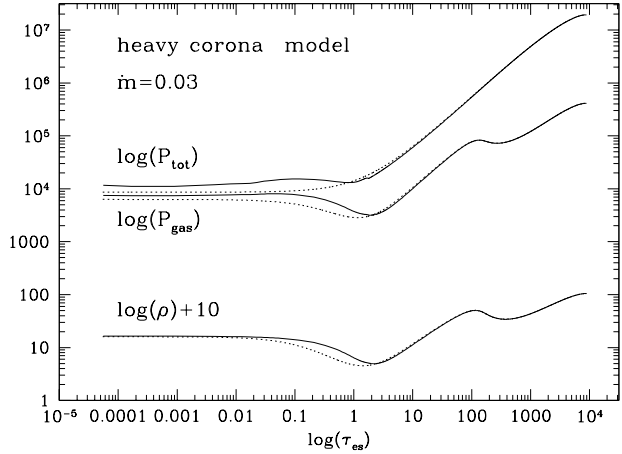}
 \caption{The distribution of density, gas pressure and total pressure 
         for the heavy corona model. Initial profiles are presented by 
dotted line, 
and final iteration by solid 
line.}
 \label{fig:gest2}        
\end{figure}

The uppermost layers are heated up only by a factor of $\sim 1.2 $
(Fig.~\ref{fig:iter3}).  
%We present only one case of corona, but one can see that though it is 
%optically thin
%the dynamical influence of the corona is large.
The weight of the corona modifies the density of the surface layers 
and we show this effect in Fig.~\ref{fig:gest2}.  
The initial high value of the density decreases towards the disc interior
forming the outer density inversion (see R\'o\.za\'nska et al. 1999).  
Such inversion is reflected in the pressure variation versus Thomson 
optical
depth (Fig.~\ref{fig:gest2}). Deeper gas density inversion is again 
connected with convection.

In the outer layers the gas pressure is comparable to the radiation 
pressure,
but towards the equatorial plane the radiation pressure dominates 
even by two orders of magnitude. 

\subsubsection{Overall spectrum}

\begin{figure}
 \epsfxsize = 75mm \epsfbox[80 200 485 500]{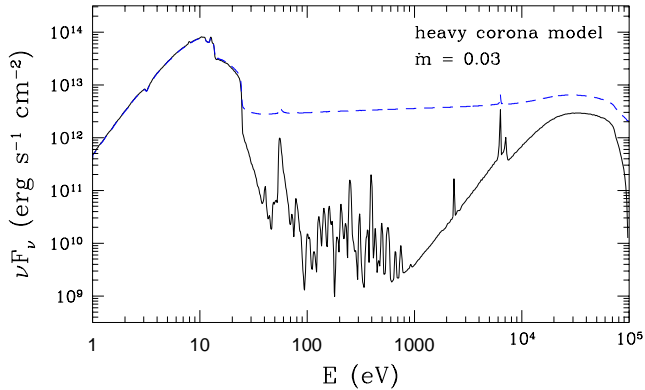}
 \caption{The spectrum of the irradiated disc with corona. 
        Reflected spectrum  - solid line, 
        Comptonized reflected component - dashed line. Accretion rate: $\dot m = 0.03$.}
 \label{fig:cspec2}        
\end{figure}

The final spectrum as seen by the observer has to be calculated taking into
account the direct presence of the corona.

In the case of a lamp post model we
 had to include, in agreement with
equation~\ref{eq:sur1}, half of the incident X-ray radiation which goes 
directly
to the observer (see final spectrum as a bold dotted line in 
Fig.~\ref{fig:spec1}).
In the case of the corona model the radiation emerged from the disc is
subsequently Comptonized by the hot coronal layer. To
describe this effect we apply a simple code of Czerny \& Zbyszewska (1991)
based on semi-analytical formulae
 appropriate for optically thin Comptonized 
medium. This code does not include the effect of the
anisotropy of Compton scattering. However, since in a bare disc model we
also simplify the description of the fraction of the primary radiation
going directly to an observer (we use the power law instead of Comptonized 
emission shape) such an approximation provides a better comparison of the
two cases. 

\begin{figure}
 \epsfxsize = 75mm \epsfbox[30 430 490 810]{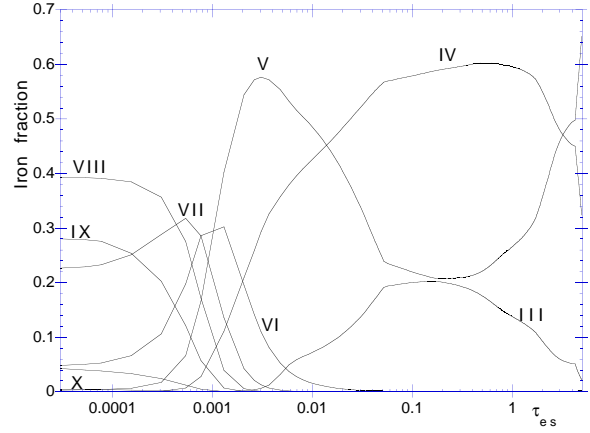}
 \caption{The fraction of abundances of iron ions for the heavy corona 
model, versus
Thomson optical depth.}
 \label{fig:cion}        
\end{figure}

\begin{figure}
 \epsfxsize = 75mm \epsfbox[30 430 490 810]{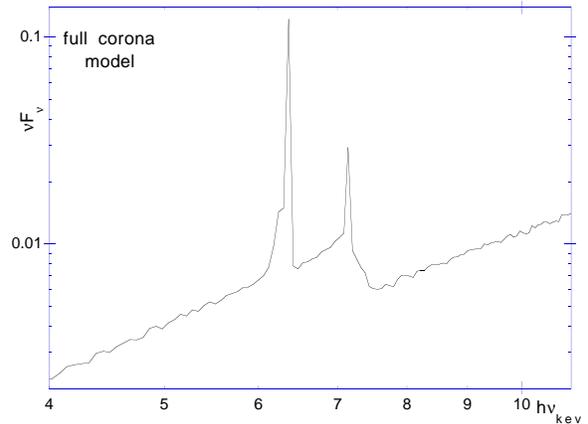}
 \caption{The close-up of the  reflected spectrum (in arbitrary units)  
in the energy band of  the  iron line, 
for the heavy corona model. Spectral resolution $R= 100$.
 Accretion rate: $\dot m = 0.03.$}
 \label{fig:clin}        
\end{figure}

In Fig.~\ref{fig:cspec2} we present the reflected local spectrum by 
continuous
 line, 
while the dashed line represents reflected spectrum modified by its 
passing  through the corona with temperature $1 \times 10^9$ K and
with $\tau_{cor} =0.355$.

Such a corona, however, cannot cover the entire disc surface.
According to our model, 
the sum of the energy emitted in X-rays and 
the energy emitted by disc cannot exceed the total energy generated due
to accretion.
Taking the assumed fraction of energy dissipated in the X-ray source  
$f=0.5$ and using Eq.~\ref{eq:sur1} (albedo equal to zero) to describe the 
total 
outgoing soft flux, the Comptonized flux is given by:
\begin{equation}
F_X= A_{C} (0.5  F_X + 0.5 F_{gen}) = A_{C} (1.5 F_X), 
\label{eq:bilan}
\end{equation}
which gives the approximate constraint on the amplification factor:
\begin{equation}
A_{C}= 0.667.
\end{equation}

Such Compton amplification in case of plane parallel coronal slab would 
give
a very steep soft spectrum, inconsistent with the slope assumed in the 
computations.  
Our Comptonization code indicates that for the parameters $\tau_{cor} 
=0.355$
and  $ T = 1 \times 10^9$ K (appropriate to reproduce the
spectral index
$\alpha =0.9$) the amplification is 3.9.
Also from simulations done by Abrassart \& Czerny (2000, their Fig.3) 
one can see that in case of a disc black body spectral distribution with
the temperature  $50$ eV 
the amplification factor never drops below $4$ for a wide range of 
coronal temperatures. This is the main reason why a uniform non-patchy
corona covering the entire disc is not a good model of Seyfert galaxies.

Since we model the entire spectrum at a single radius, we
can circumvent this problem by breaking down the assumption of 
full, continuous  corona and consider the case of a  corona covering the 
disc only 
partially. We introduce a new parameter, $f_{geom}$, which decreases the
probability of scattering of the soft photons in the corona due to its 
patchy
structure. This parameter has to be
included in Eq.~\ref{eq:bilan} as:
\begin{equation}
F_X= A_{C} f_{geom}(0.5  F_X + 0.5 F_{gen}) = A_{C} f_{geom} (1.5 F_X), 
\label{eq:bil2}
\end{equation}
and implies the amplification factor to be:
\begin{equation}
A_{C}= \frac{0.667}{f_{geom}} .
\end{equation}
Varying $f_{geom}$ we can find the solution which preserves the required 
spectral slope, and the amount of coronal material which has
to cover the cold disc to maintain energy balance between the hot and cold 
phases.

Fig.~\ref{fig:cspec2} shows that the spectrum reflected from the disc part 
covered by the corona and the 
observed spectrum from that radius
 is clearly a reflection from a basically neutral gas. No strong
soft X-ray lines, characteristic of the lamp post geometry, are visible.
Even at the
disc surface the iron is only weakly ionized (Fig.~\ref{fig:cion}). Only 
the
6.4 keV iron line is visible in the expanded picture of the 4-10 keV band,
with relatively strong K$_{\beta}$ line and considerable edge 
(Fig.~\ref{fig:clin}).

However, the final spectrum of the object is not just a spectrum at the 
radius covered by the corona, but also the contribution from the 
uncovered part has to
be included, in agreement with the required covering fraction.
In our specific case with parameters: spectral slope $\alpha = 0.9$ 
and 
fraction of energy
dissipated in the corona equal 1/2, $f_{geom}$ is very
small, $\sim 0.13$, and the additional contribution ($ \sim 0.87$)
 from disc radii uncovered
by the corona will dominate the total spectrum. In this case it would be
impossible to distinguish between the lamp post and the patchy corona 
models.

However, in sources with a weaker Big Blue Bump (i.e. a 
small fraction of energy
dissipated in the disc) $f_{geom}$ is about 0.5 and the spectrum from
a heavy corona model would approximately look like the spectrum in the 
 lamp post geometry but with soft X-ray lines suppressed by a factor 
of order of 2.  
In sources
with softer (steeper) X-ray spectra, slab amplification is significantly 
lower
and the clumpiness of the corona is not needed. Therefore, in soft spectra
sources, whenever the disc is covered by the corona, we would not expect 
any 
soft X-ray lines.

\subsection{Comparison with observations}

The are several interesting new observations of iron line done by
satellites as BeppoSAX, XMM and CHANDRA. They usually report that iron
line is rather narrow without relativistic broadening (Yaqoob, George, 
Nandra et al. 2001). 
Reeves et al. (2001)
showed observations of the radio quiet quasar Markarian 205 where 
two components of iron line are present: a narrow neutral line at 6.4 keV
and a broadened line centered at 6.7 keV.
They concluded that those observations are consistent with a disc origin 
only
if the disc ionization is high enough to produce He-like iron, and if the 
narrow neutral component is produced elsewhere.
But those conclusions were obtained by fitting  the
data with constant density models. 
Our model of illuminated disc in hydrostatic
equilibrium shows that  both components of iron line are present.
The ionization state
varies across the atmosphere and therefore we get a recombination
iron line from uppermost layers and a fluorescence line from
deeper zones. Also the EW of fluorescence line in
our model ($EW=40$ eV for $\dot m = 0.03$) 
is not in bad agreement with the observations
(see Reeves et al. 2001 Table 1) where EW is 46 eV. 

Another interesting observation, which could be
fitted by our model,  was reported by Comastri, Stirpe, Vignali 
et al. (2001). They observed the bright Narrow-Line Seyfert 1 galaxy Ark 
564 and
they reported a strong iron line from He-like iron  line at  $\sim 6.76$ 
keV.
The EW of the line estimated from an ionized reflection model
is $\sim 87$ eV. 
In our model with high accretion rate we obtain predominantly 6.7 keV
component, with EW of $\sim 90 $ eV. 
However, we must notice that this source has much softer
X-ray spectrum (photon index $\Gamma \sim 2.5$, Comastri et al. 2001,
Turner et al. 2001). 

Using our model, the emission lines from other elements such as oxygen, 
nitrogen and  carbon can be determined.
Branduardi-Raymont et al (2001) have observed the spectrum of 
MCG -6 -30 -15 in the soft X-ray range with XMM. They claimed that 
the very intense and broad features in this object cannot be 
attributed to a warm absorber, but are emission  lines produced by an 
irradiated accretion disc and relativistically broadened.
This result was not confirmed by the analysis of the CHANDRA data,
as reported by Lee et al. (2001).
Our model shows that irradiated accretion discs with moderate illumination {\it cannot} produce 
very strong emission lines. The results presented in this paper  
(Figs.~\ref{fig:spec1}
and \ref{fig:spechigh}) 
show that the most prominent line
(doublet  Ne$_{X}$ Fe$_{XVII}$) has $EW \sim 18$ eV (for both 
$\dot m = 0.03$ and $\dot m = 0.3$) in comparison to the 
emitted continuum.

\section{Discussion}

\subsection{Precision of the method, advantages and drawbacks}

\subsubsection{Radiative transfer}

In the present paper we do not recall the influence of the
transfer treatment, which was discussed in DAC 2000 and in Dumont \&
Collin 2001. It was for instance shown that it is most important to take
into account
correctly the transfer of the diffuse continuum (which is not performed in
Ballantyne et al. 2001), and that the escape
probability approximation gives incorrect results
for thick inhomogeneous slabs. Not only this approximation can lead to 
differences in the 
line intensities, but also to differences in the energy balance, which is 
dominated by the line heating/cooling in a large fraction 
of the illuminated layer. 
The importance of these processes will be discussed in more
detail in a forthcoming paper devoted to this issue 
(Coup\'e et al. 2002).  Their correct description is clearly 
an advantage of our model.

\subsubsection{Atomic data}

The importance of heavy elements in illuminated disc atmospheres was
clearly shown by Madej \& R\'o\.za\'nska (2000).
Thermal absorption is completely inefficient in pure hydrogen-helium 
atmospheres, 
and incident X-ray radiation is redistributed by Comptonization.
Iron rich models considered by Madej \& R\'o\.za\'nska (2000) include
only bound-free absorption on iron, but they show completely different 
spectra 
with smaller UV bump and a lot of ionization edges 
which do not exist when Iron is not taken into account.

\begin{figure}
 \epsfxsize = 70mm \epsfbox[45 20 660 520]{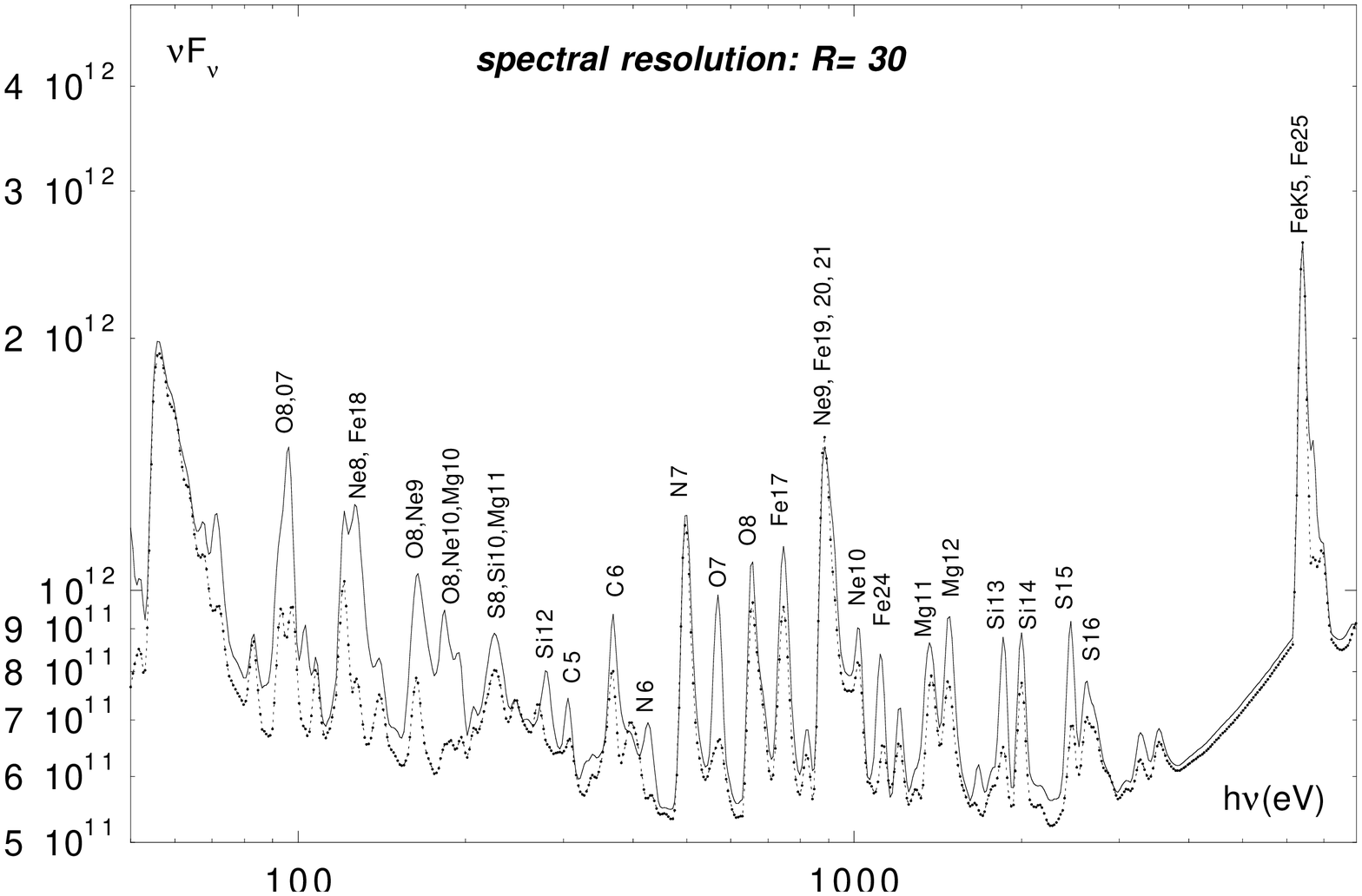}
 \caption{Influence of the atomic data on the reflection spectrum for the 
 lamp post model with $\dot{m}=0.03$. Solid line: multi level Li-like, He-like, 
 and H-like ions, dashed line: only multi level H-like ions. 
Significant difference is only seen below 200 eV.}
 \label{fig-atom-data}        
\end{figure}

Obviously  the atomic models and atomic data determine directly the line 
intensities. However 
they have also an influence on the fractional ionic abundances, through 
the ionization from, and the recombination onto, the excited states.
Indeed close to LTE, i.e. in the dense
layers of the atmosphere, ionizations from excited levels almost
balance recombinations onto these levels, while in the hot skin, only
recombinations onto excited levels are important. Improved atomic
data have been recently implemented in the code (cf. Coup\'e et al. 
2002). In particular all Li-like and He-like ions are now treated in a 
complete interlocked way with 5 to 9 levels plus the continuum, while in 
the old version of {\sc titan} only H-like atoms were treated with several 
interlocked 
excited levels. This is still
 less than in 
XSTAR, used by NKK00, but much more than Ballantyne et al. (2001). These new 
data have been used to compute the spectrum
 in the lamp post moderate accretion rate 
case, and they are shown
  to have an
impact on the spectral features in the EUV,
 but basically not on the vertical disc
structure and on the overall continuum spectral distribution (cf. 
Fig. \ref{fig-atom-data}).

\subsubsection{Thermal instability issue}
\label{sect:instab}

Another problem in which one has to choose an approximation is in the 
approach
of the thermal instability.

The vertical structure of the surface irradiated layers of  accretion discs
is very complex (R\'o\.za\'nska \& Czerny 1996, hereafter RC96). 
Illumination of gas in hydrostatic equilibrium
by hard X-rays implies the presence of a thermal instability. More
precisely, in certain pressure range, the gas can be in three 
states of thermal equilibria, with an intermediate one being unstable.
The problem is best seen as the presence of S-shape feature in the 
temperature
versus the ionization parameter $\Xi$ plot (Krolik, McKee \& Tarter 1981).
There is no unique solutions for
the temperature and density profile. Physically, we expect a 
transition from the upper stable Compton-heated branch to the lower stable 
branch determined by atomic processes. This transition must happen 
somewhere
in the multi-solution zone and it may be sharp or a two-phase medium may be
present.

Unique strict solution of the vertical structure can only be found if
electron conduction, in addition to the radiative transfer, is included 
(R\'o\.za\'nska 1999). However, conduction complicates the numerical 
scheme 
considerably and at present it was never combined with computations
of radiative transfer. In R\'o\.za\'nska (1999) radiative transfer was 
replaced by the energy balance 
equation, where heating and cooling functions were determined by means of 
photoionization code {\sc cloudy}  appropriate only for moderately thick 
media 
(DAC00).

Other papers neglected the conduction term but contained more advanced 
radiative transfer computations. These papers can be divided into two 
families,
depending on the way of iterating between hydrostatic equilibrium and
radiative transfer, which implies
their approach to the thermal instability problem.

First family solves strictly both the radiative transfer and hydrostatic 
equilibrium, iterating density while solving radiative transfer. In this 
case the problem of appearing multiple solutions is solved by an arbitrary
 choice of the transition pressure and introducing a discontinuity. This 
discontinuity, chosen
 in the first iteration, is preserved in later computations. This approach 
was taken by Ko \& Kallman 1994, and subsequently
by Sincell \& Krolik (1997, 1999) and Nayakshin, Kazanas \& Kallman (2000,
hereafter NKK00). The choice of the position of this discontinuity 
does not influence strongly the resulting spectra 
(Jimenez-Garate et al. 2001). 

Second family keeps the density constant while iterating the 
temperature at a given computation step. Such a numerical scheme always 
produces a unique 
solution across the transition region but at the expense of not providing 
a strict solution for both radiative transfer  and hydrostatic 
equilibrium
in illuminated atmosphere.
This approach was used by Raymond (1993), and subsequently by Shimura, 
Mineshige \& Takahara (1995), Madej \&  R\'o\.za\'nska (2000), 
Kawaguchi, Shimura \& Mineshige (2001), and Ballantyne, Ross \& Fabian 
(2001).

Both approaches provide an approximate answer to the problem since the 
first one does not effectively iterate the optical depth of the hot zone 
adopted during the first iteration while the second one is never fully 
converged
in the transition zone. 

\begin{figure}
 \epsfxsize = 80mm \epsfbox[70 20 750 520]{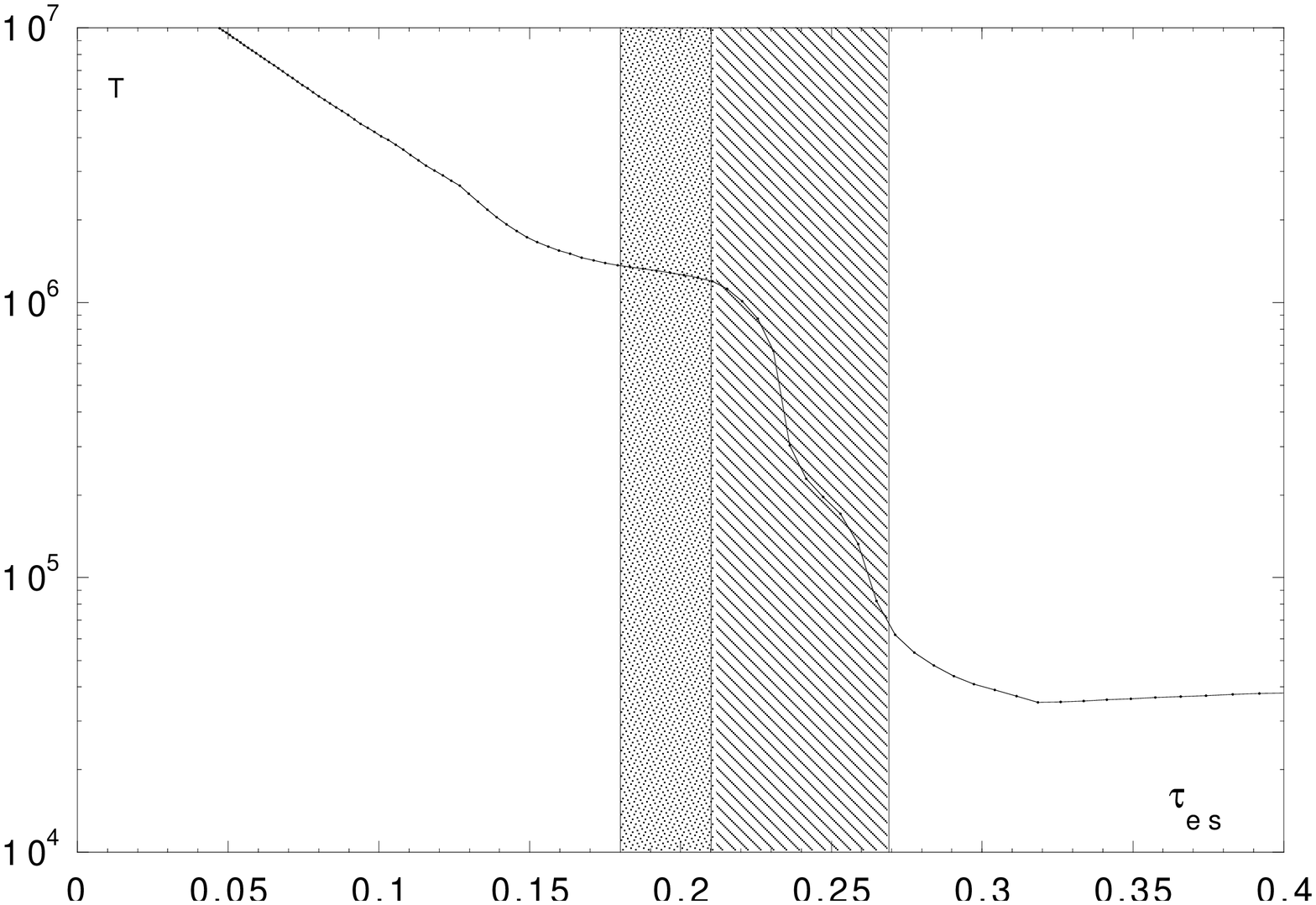}
 \caption{The temperature profile for the 
   lamp post model computed with $\dot{m}=0.03$. Grey region: the
 multiple solution region where the hot solution is chosen; hatched 
 region: 
 ill defined solution (see text).}
 \label{fig-expanded-T}        
\end{figure}

In the present paper we have adopted the second approach. This method gives a
solution for the equilibrium temperature which corresponds to an ill 
defined
value of the pressure. The  extension of this layer is indicated 
in Fig.~\ref{fig-expanded-T} 
which is an expanded version of Fig. \ref{fig:iter},
 together with the extension 
of the multiple solution region. We see that the 
 ill-defined region is relatively thin, so one would expect that it has no
 influence on the overall emitted
spectral distribution, but still it may lead to uncertainties in the detailed
 spectral 
features, anyway inherent to the presence of the thermal instability;
 in particular, it contains several important iron ions such as
 FeXVII. On the other 
hand the hot solution is suppressed at $\tau_{es}=0.21$, whilst the 
transition to the cold zone should 
probably occur at a slightly smaller optical thickness, when conduction
 is taken 
into account  (cf. below). 
We have also checked that increasing the number of layers by a factor 2 
in the transition zone does not change at all its structure.

\subsubsection{Expected effect of conduction}

Since our current approach does not include yet the conduction term we
test the possible effect of such a term using the local approach of
R\'o\.za\'nska (1999) and computing the solution for $\dot m = 0.03$ with
and without thermal conduction.

Apart from the surface temperature given by the inverse Compton value, one 
more 
boundary condition was required to solve the second order differential 
equation of the energy balance with conduction. This condition was chosen 
as the
requirement that at the bottom of the computed zone the solution matched
that of the disc: when the same pressure as in the previous computations 
was reached,
the temperature was also the same.

\begin{figure}
 \epsfxsize = 80mm \epsfbox[25 300 550 700]{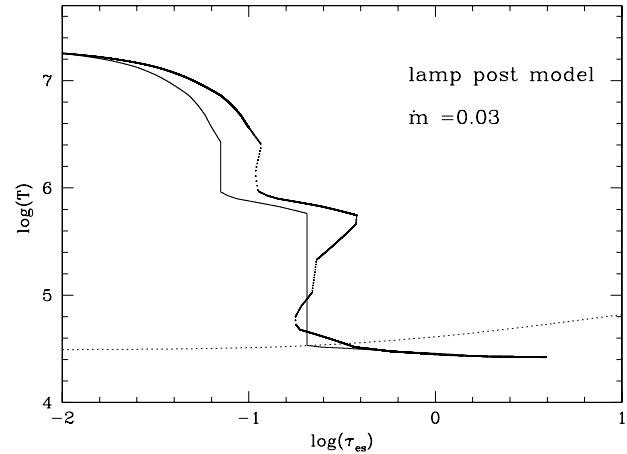}
 \caption{The temperature profile for the 
   lamp post model computed with a local approach  with
(solid line) and
  without (thick points) thermal conduction.
  All profiles are computed for the same parameters as the lamp post model 
  with $\dot{m}=0.03$. Dotted line has the same meaning as in Fig.~\ref{fig:iter}.}
 \label{fig:iter4}        
\end{figure}

The dashed line in Fig.~\ref{fig:iter4} represents the temperature profile 
obtained without radiative transfer but from cooling and heating functions
determined from CLOUDY (for better description see  RC96).
The solid line shows the same computations
taking into account thermal conduction. 
The disc flux, the X-ray flux and the spectral distribution of
illuminating radiation are the same as in the lamp post model with 
$\dot{m}=0.03$.

Clearly, the approach based on local cooling cannot asymptotically
reproduce the temperature profile given by the 
diffusion approximation. The local approach indicates also  a certain 
flattening
 just below $ 10^6$ K 
which is
less important in the {\sc titan} solution. 
This is due to the lack of the true radiative transfer in this simple exercise.

However we see that in the upper hot zone, the temperature drop 
occurs faster when thermal conduction is taken into account, but finally 
the total optical depth of the Compton heated zone 
with thermal conduction 
is not changed considerably. 
Therefore, we conclude that the optical 
depth of the hot layer is determined accurately enough in our radiative 
transfer calculations without thermal conduction.

\subsection{Comparison with other methods}

There were many computations of the vertical structure and of the 
spectrum of irradiated discs for AGN and X-ray binaries in the two past 
decades.  
The most recent ones which present spectra including lines from illuminated 
disc 
atmosphere with heavy elements in the case of hydrostatic equilibrium
are NKK00 and Nayakshin \& Kallman (2001), and Ballantyne et al. (2001).

It is extremely difficult to compare directly our results to them,
 as our computations were performed before the publication 
of these papers, and therefore adopt other values of the various 
parameters. 
For instance the 
spectral distribution of the incident spectrum is different, a fact of most 
importance for the structure of the irradiated skin. We have 
mentioned in the introduction that the spectrum of an irradiated thick layer
 has been computed
using the three codes with exactly the same parameters in the case of a 
constant density medium  (cf. Pequignot et al. 2001). 
The three computed spectra differ in their 
overall spectral distribution and in their
detailed features. Concerning the model of irradiated disc in 
hydrostatic equilibrium, for the moment the only possible comparison
 is to stress the differences between the methods which are used. 

Ballantyne et al. results are limited 
to very high fluxes by their computational method, because their 
code has difficulty if the hot skin has $\tau_T \le 0.2$ (see Sec.6 
Ballantyne et al. 2001).  
%so they assume $F_{X}/F_{disc} > 10 $.
  It precludes any 
comparison with their results in the present paper, but this will be 
achieved in a following paper, dealing with high illumination. 

NKK00 use for the thermal and ionization equilibrium 
the code XSTAR. The advantage is that
 the atomic 
physics is better taken into account than by us, but the drawback
is that the 
total energy balance is not correctly determined (cf. DAC2000 an Coup\'e 
et al. 2002). The local energy balance is also not well determined in some 
cases. In particular, NKK00 were obliged to set the gas 
temperature equal to the effective one when it was lower.
 This 
prevents them to compute the temperature in the deepest irradiated 
layers.
 It means that above a given optical thickness, which is of 
the order of 0.2 in a case comparable to our lamp post models, the 
temperature becomes constant. This is different from the real 
temperature structure (cf. Fig.~\ref{fig:iter}).
Moreover, they do not compute like us the vertical structure of the 
underlying disc, but they start their integration from a  height 
given by the Shakura-Sunayev (1973) vertically averaged solution. We 
show analytically in the appendix that this is valid only if the disc 
scale height is negligible with respect to the scale height of the 
irradiated skin, i.e. for low accretion rates ($< $ 0.01 in Eddington 
units). We have also a completely different way of handling the thermal 
instability,
as NKK00 simply stop their computation when the hot solution disappears, 
while we compute the structure in the region where the 
cold solution applies.

The same problems occur in the paper of Ballantyne et al. (2001), 
where the basis of the hot skin
is fixed at the vertically averaged half-thickness of a gas-pressure 
dominated disc. 
Also those authors use poor description of heavy elements with
only few ionization stages and excited levels.
 This treatment should underestimate the 
photoionization
heating and the line cooling. 

The transfer of the continuum is performed basically in the
same way in NKK00 and in our computations, which is 
better than the approach of 
Ballantyne et al. who apply the diffusion approximation to the transfer of the 
diffuse radiation (Ross \& Fabian 1993). However both NKK00 and Ballantyne 
use the Companeets equation for the Compton 
diffusions, while we take the inelastic scattering 
into account only above 1 keV, through 
the coupling of a Monte Carlo code. This is clearly a disadvantage of our 
method. The transfer of the
lines differ also strongly between us and them.
 Above 1 keV, we use the Monte Carlo code which
takes into account Compton
diffusion, and below 1 keV, we perform real transfer
computation in the lines, but we do not take into account
comptonization. So we have the 
advantage of performing real transfer in the lines, but the 
disadvantage of not 
taking into account Compton diffusions.
This can have some impact on the intensities of
resonance lines which are subject to a large number of scatterings. All other
authors use the escape probability formalism for the lines, which however 
has the advantage to take into account the escape by Compton diffusions.

In their lamp post computations, 
 Nayakshin \& Kallman (2001) assume a smaller X-ray flux, with the luminosity 
of the X-ray source being 
 20$\%$ of the integrated disc luminosity, but nevertheless it  corresponds to a 
larger local flux than ours
(they compute the spectrum at 6$R_{Schw}$ and not at 
 10$R_{Schw}$). 

 NKK00 introduce the ``gravity parameter" $A$ which is the ratio of 
the vertical component of the gravity force at one disc scale height to the 
radiation pressure force 
(for the definition, see NKK00 Sect. 3.3), and they give their results as 
functions of this parameter. 
If we calculate the value of  parameter $A$ of NKK00 from our computations of the
disc vertical structure we obtain  
$A\sim 2$  for both accretion rates. Indeed this parameter is not 
sufficient to determine the solution, one must also define the X-ray flux 
since $A$ is proportional to the ratio of 
the viscous to the X-ray flux.  So in
the following paper, Nayakshin \& Kallman (2001) present the NKK00 lamp post
 models for different
 accretion
 rates and X-ray fluxes. Our two models have about the same ratios of 
 viscous to X-ray fluxes and 
about the same accretion rates as their models w4 and 
 w7. Our models
predict an optical depth of the entire hot layer of about 0.2 and 0.4, 
while in Nayakshin \& Kallman the hot layer is almost twice larger. This 
can be partly due to the spectral distribution of the incident continuum. 
We adopt a power law with a slope of 0.9 with a cut-off at 100 keV, while 
their power low has a slope of 0.8 and the continuum
curves slowly above 100 keV due to the adopted cut-off. We see clearly 
that a direct comparison is not possible.

Finally the effect of coronal influence is studied only in case
of purely hydrogen atmosphere in the paper of 
Kawaguchi, Shimura \& Mineshige (2001). Since they do not take into 
account  
heavy elements, the disc does not absorb 
X-rays and produces relatively less seed photons for Comptonization.
It leads to less efficient coronal cooling and produces rather hard X-ray 
spectrum even if corona covers the whole disc.
This is not the case in our calculations. The disc atmosphere with
heavy elements has a low albedo and all absorbed X-ray photons are 
reemitted in the soft band as  seed photons for Comptonization.
Therefore, to produce hard X-ray spectrum the corona has to cover
the disc only partially.

\section{Conclusions}

In this paper we have presented computations of the vertical structure and 
spectrum of an  
illuminated accretion disc, using a new code available for Compton thick 
media,
{\sc titan/noar}, and taking into account hydrostatic equilibrium. 
The disc atmosphere is integrated
vertically together with the disc interior.
Below the illuminated atmosphere, for
$\tau_{es} >6$, diffusion approximation is used. 
We compare two geometries: the lamp post model, where an X-ray source is 
situated above the disc at some distance, and the heavy corona model, where
the corona, besides illumination, exerts a dynamical pressure on the disc 
material.  
We found:
\begin{itemize}  

\item In the case of the lamp post model, the uppermost layers of 
atmosphere 
are heated to very high temperature
which drops suddenly with increasing optical depth. Such a sharp 
transition is
consistent with theoretical predictions and, depending on strength of 
illumination, may indicate thermal instabilities (KMT81, RC96).
The radiative transfer computations prevent the temperature 
profile to be multivalue, nevertheless thermal instabilities occur as
a 'wiggle' in the pressure profile.
To avoid this problem thermal conduction should be taken into account and
we plan to do it in the future work.
%The lack of those instabilities in computations done by other groups may 
%indicate an underestimation of heating and cooling processes, mostly
%photoionization and cooling in lines (Ballantyne et al. 2001). 

\item The disc scale height is important for accretion rates $\dot m 
>0.01$ 
(see appendix) and it influences the optical thickness of the illuminated 
zone.
Therefore, it is essential to integrate carefully the cold region below
the transition, in order  to determine the fractional abundances of weakly 
ionized heavy elements, which influence the extreme UV and soft X-ray 
spectrum.

\item In case of the lamp post model, the outgoing spectrum
differs in extreme UV and soft X-ray band from the case of
constant density slab, indicating, that 
it is extremely important to estimate the temperature and the density in 
the partially ionized medium. Also K$_\alpha$ iron line looks different in 
both cases.
The absorption
edge is absent contrary to the constant density case.

\item A heavy corona, even optically thin, completely suppresses the 
highly ionized zone on the top of the accretion disc.

\item The spectrum from the disc covered by a heavy corona is featureless,
but it depends on the expected energy distribution of incident 
radiation. When the X-ray spectrum is hard the energy balance requires that
corona covers the disc only
partially, and the spectrum  from the non covered part becomes important.
In such a case it is impossible to distinguish the presence or absence of
a patchy corona. 
When the X-ray spectrum is soft, we can infer the presence of a corona
in those objects with featureless spectra.  

\item Our model is in agreement with recent observations of  K$_{\alpha}$ 
iron
line, where in some objects only the fluorescence component is present 
(NGC 5548 Yaqoob et al. 2001), in others, only the recombination component 
is present 
(Ark 564 Comastri et al. 2001), or like in Mkr 205 (Reeves et al. 2001)
both components are seen. It can be explained in the framework of the  
our model by changing the illuminating X-ray flux. 

\end{itemize}

\appendix
\section{Semi-analytical results}
\label{sec:serg}

A crucial question is how the optical depth of X-ray heated zone 
depends on explicit  calculations of the disc vertical structure.

Let us consider hydrostatic equilibrium with  gravity $g$ and 
ignoring  the term of radiative pressure.
We can then simply calculate the optical depth of the hot zone as:
\begin{equation}
\tau_{es}^{hot}= \frac{P_{bot} \kappa_{es} }{g},
\end{equation} 
where $P_{bot}$ is the gas pressure on the bottom of the X-ray heated skin,
which can be expressed as:
\begin{equation} 
P_{bot} = \frac{F_X }{\Xi_{bot} c}.
\end{equation} 
The ionization parameter on the bottom, 
$ \Xi_{bot}$, is equal to the $ 1.22 (T_{IC}/10^8)^{-3/2}$ 
(Begelman, McKee, Shields 1983), with the 
inverse Compton temperature $T_{IC}$ defined as  usual. 

In the case of stellar atmospheres, the optical depth of X-ray heated 
layer is a function of only the hard X-ray flux and the shape of incident 
radiation
(through the $T_{IC}$):
\begin{equation} 
\tau_{es}^{hot}= \frac{F_X \kappa_{es}}{ \Xi_{bot} \, c \, g }.
\end{equation} 

In the case of an accretion disc, the gravity depends on the distance from 
the 
black hole and on the pressure scale-height of the considered layer 
(we  assume the gravity to be vertically constant). 
Considering pressure scale-height of the full system i.e. the illuminated  
zone plus the disc
$H_p=H_{hot}+H_d$, we can simply derive:
\begin{equation} 
\tau_{es}^{hot}=\frac{F_X \kappa_{es}}{ \Xi_{bot} \, c \, \Omega_K^2} 
   \left( \frac {1}{H_{hot} + H_d} \right),
\label{eq:skin} 
\end{equation}   
where $H_{hot}$ is the scale-height of the isothermal Compton heated skin.
If $ H_{hot} $  is  much larger than $H_d$ the  disc
scale-height can be neglected.    
Rewriting  Eq.~\ref{eq:skin} with  
Eq.~\ref{eq:schei} for constant gravity case and replacing $T_{cor}$ by 
$T_{IC}$ one can get:
\begin{equation}
\tau_{es}^{hot} = \frac{F_X \kappa_{es}}{ \Xi_{bot} \, c \, \Omega_K} 
   \left( \frac {\mu m_H}{ k_B T_{IC} } \right)^{1/2} .   
\end{equation} 
 
%i.e. $H_d=0.23 \, \dot m R_{Schw} 
%(1-\sqrt{3/R} ) $, where  $R=r/R_{Schw}$ (Kato, Fukue, Mineshige 1998, 
%p.87) 
%we obtain:
%\begin{eqnarray}
%\frac{H_{hot}}{H_d} & = &\frac {1}{0.23} \left(\frac{2 \, k_B}{ \mu m_H 
%c^2} \right)^{1/2}
%  \frac {R^{3/2} T_{IC}^{1/2}} {\dot m (1-\sqrt{ 3 / R} )} \\
%  & = & 2.63 \times 10^{-6}
%  \frac {R^{3/2} T_{IC}^{1/2}} {\dot m (1-\sqrt{ 3 / R} )}.
%\end{eqnarray}
%Note, that the ratio above does not depend on the mass of the central 
%object, therefore
%it is the same for the case of AGN and galactic black holes. 

For the case of variable gravity the hydrostatic equilibrium should be 
solved
together  with the equation of mass continuity $d\tau= \kappa_{es}  \rho 
dz $.
$\tau_{es}^{hot}$ is then larger by a factor of  $\sqrt{2}$ 
(the same factor as for the pressure scale-height, see Eq.~\ref{eq:schei}),
and by integral of order of unity (Nayakshin 2000, Eq.2). 

\begin{figure}             
 \epsfxsize = 80mm \epsfbox[25 300 550 700]{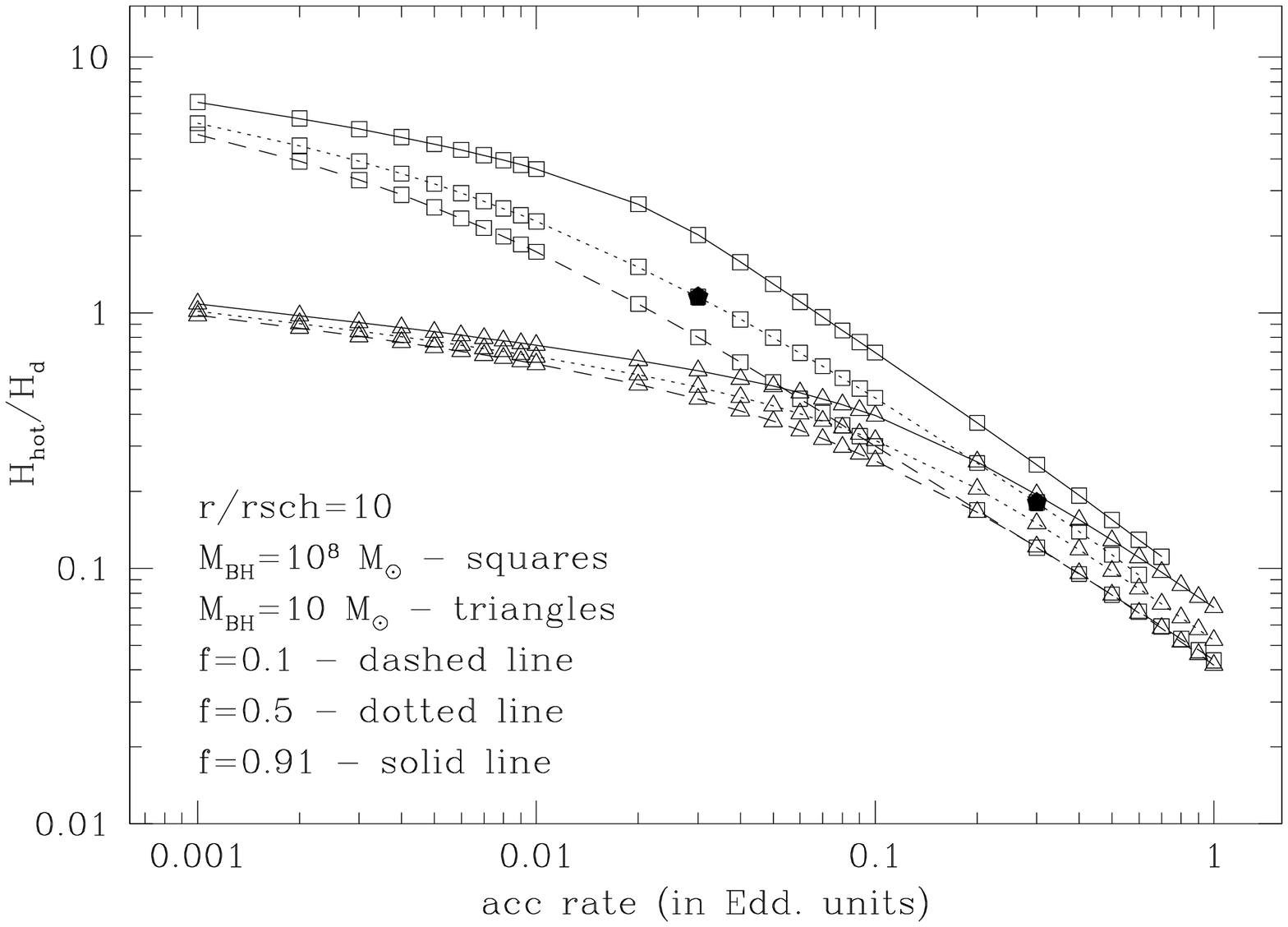}
 \caption{ The ratio of scale-height of isothermal Compton heated skin to 
the 
          disc scale-height, versus the accretion rate in Eddington units,
          for a given radius ($r=10 R_{Schw}$) and for different fractions 
of energy 
          dissipated in the X-ray source:
         $f=0.1$ - dashed line, $f=0.5$ - dotted line, and  $f=0.91$ - 
solid line.
         The results are presented for two different masses of the black 
hole:
         $M_{BH}=10^8 M_{\odot}$ - squares, and $M_{BH}=10 M_{\odot}$ - 
triangles.
         Filled squares mark the two cases computed numerically and
considered in detail in the previous 
sections
         $\dot m =0.03$ and $\dot m =0.3$.}
\label{fig:serg}        
\end{figure}

If $ H_{hot} $ is comparable to  $H_d$, the disc structure is important 
and the optical depth of the X-ray heated skin cannot be determined 
without 
proper disc calculations.
Fig.~\ref{fig:serg} gives the ratio  $H_{hot}/H_d$  versus accretion rate 
for the representative radius $r=10 R_{Schw}$. $H_d$ is computed 
by solving the explicit disc vertical structure (see 
Sec.~\ref{sec_firststep}) 
instead of assuming vertically averaged Shakura \& Sunayev disc (1973) 
like in NKK00.
The computations are done for two black hole masses: 
$10^8 M_{\odot}$, which corresponds to the AGN, and  $10 M_{\odot}$, 
corresponds to GBHs.
We assume different values of $f$ from low illumination $f=0.1$, which 
corresponds to $ F_{X}/F_{disc} = 0.11$, to strong illuminaton 
$f=0.91$ which gives $ F_{X}/F_{disc} = 10 $.

\begin{figure}
 \epsfxsize = 80mm \epsfbox[30 400 550 700]{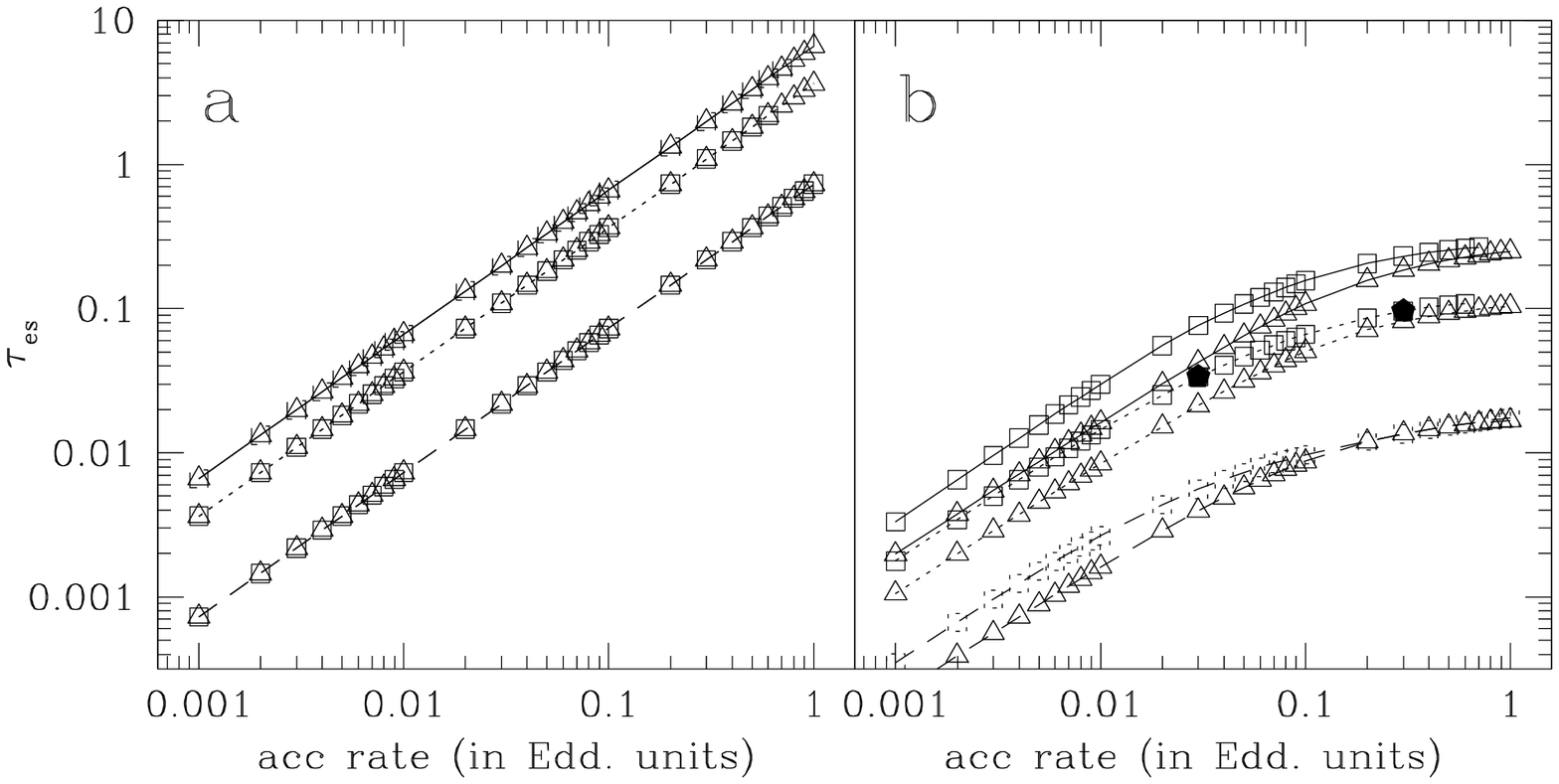}
 \caption{The dependence of optical depth of X-ray heated skin on the
         top of accretion disc  on the 
        accretion rate in Eddington units, for the  given radius ($r=10 
R_{Schw}$) 
       and for different fraction of energy dissipated in the X-ray source:
       $f=0.1$ - dashed line, $f=0.5$ - dotted line, and  $f=0.91$ - solid 
line.
       Results are presented for two different masses of the black hole:
       $M_{BH}=10^8 M_{\odot}$ - squares, and $M_{BH}=10 M_{\odot}$ - 
triangles.
       Panel a represents results computed according to Nayakshin (2000) 
assuming 
       integral equal 1, and panel b is computed for constant gravity 
Eq.~\ref{eq:skin}
        but including the influence of disc structure.
         Filled squares mark the two solutions computed numerically and 
considered in the previous 
sections
         $\dot m =0.03$ and $\dot m =0.3$.}
 \label{fig:stau}        
\end{figure}

One can see that the ratio is slightly higher for higher $F_X$ and
increases with decreasing accretion rate.
Note that in case of GBHs the ratio is always smaller than 
or equal to unity implying that
the disc structure is important for all accretion rates. 
In case of AGN for $\dot m > 0.04 $ the vertical structure of the disc 
should 
be integrated down to the 
midplane to determine correctly the optical depth of the hot skin.

Fig.~\ref{fig:stau} presents the Thomson thickness of the 
X-ray heated skin versus  the accretion rate for
the radius $10 R_{Schw}$ which is expected to be the region where
the iron line is produced. 
We compare the analytical results of Nayakshin (2000), who 
neglect the disc scale-height and shown that this is in agreement
with NKK00 numerical computation (see Nayakshin 2000 Fig.1)
with ours Eq.~\ref{eq:skin}. 
One can see that taking into account proper disc calculations it is
impossible
to obtain  optically thick illuminated skin even for the case when 
$ F_{X}/F_{disc} = 10 $.  Therefore, when observations imply that the
hot layer is optically thick, it may be evidence for the
existence of hot corona.

\section*{Acknowledgements}

This work was supported in part 
by grant 2 P03D 018 16 of the Polish State Committee for 
Scientific Research and by Jumelage/CNRS No. 16 
``Astronomie France/Pologne''.

\ \\
This paper has been processed by the authors using the Blackwell
Scientific Publications \LaTeX\  style file.


\bigskip

\bigskip


\begin{thebibliography}{}

\bibitem []{} Abrassart A., Czenry, B., 2000, A\&A, 356, 475

\bibitem{} Alexander, D.R., Johnson, H.R., Rypma, R.L., 1983, ApJ, 272, 773

\bibitem{} Allen, C.W., 1973, Astrophysical quantities. University of 
London,
The Athlone Press

\bibitem{} Ballantyne, D.R., Ross, R.R., Fabian, A.C., 2001, MNRAS, 327, 10 

\bibitem{} Begelman, M.C., McKee, C.F., Shields, G.A., 1983, ApJ, 271, 70

\bibitem []{} Branduardi-Raymont, G., Sako, M., Kahn, S. M., Brinkman, A. 
C.,
Kaastra, J. S., Page, M. J., 2001, A\&A, 365, L140

%\bibitem []{} Brunner, H., Mueller, C., Friedrich, P., Doerrer, T., 
%Staubert, R., Riffert, H., 1997, A\&A, 326, 885 AGN - data, models

\bibitem []{} Collin, S., Abrassart, A., Czerny, B., Dumont, A.-M., 
Mouchet, M., 2000, to appear in proc. "AGN in their Cosmic Environment", 
Eds. B. Rocca-Volmerange \& H. Sol, EDPS Conf. Series in Astron. \& 
Astrophysics (astro-ph/0003108)

\bibitem []{} Comastri, A., Stirpe, G.M., Vignali, C.,
 Brandt, W.N., Leighly, K.M., Fiore, F., Guainazzi, M., Matt, G., 
Nicastro, F., 
 Puchnarewicz, E. M., Siemiginowska, A., 2001, A\&A, 365, 400

\bibitem []{} Coup\' e S., et al., 2002, in preparation

\bibitem []{} Czerny B., Zbyszewska M., 1991, MNRAS, 249, 634

\bibitem []{} Dumont A.-M., Abrassart A., Collin S., 2000, A\&A, 357, 823 
[DAC00]

\bibitem []{} Dumont A.-M., Collin S., 2001, in ``Spectroscopic 
Challenges of Photonized Plasma'', Eds. G. Ferland \& D. Savin, 
ASP Conference Series Vol. XXX

\bibitem []{} Dumont, A.-M., Czerny, B., Collin, S., Mouchet M., Zycki, 
P., 2001 in preparation


\bibitem []{} Haardt, F., Maraschi, L., 1991, ApJ, 380, L51

\bibitem[]{} Hubeny, I. 1990, Structure and Emission Properties of
   Accretion Discs, IAU Colloq. 129, 227, Eds Bertout, C., Collin, S., 
Lasota, J.-P. 

\bibitem[]{} Jimenez-Garate M.A., Raymond J., Liedahl D.A., Hailey C.J., 
2001, ApJ, 558, 448

\bibitem []{} Kawaguchi, T., Shimura, T., Mineshige, S., 2001, ApJ, 546, 
966

\bibitem []{} Ko, Y.-K., Kallman,T.R., 1994, ApJ, 431, 273 

\bibitem []{} Koratkar, A., Blaes, O., 1999, PASJ, 50, 559

\bibitem []{} Krolik, J.H., McKee, C.F., Tarter, C.B., 1981, ApJ, 249, 422 
[KMT81]

\bibitem []{} Lee, J.C., Ogle, P.M., Canizares, C.R., Marshall, H.L., 
Schulz N.S.,  
Morales, R., Fabian, A.C., Iwasawa, K., 2001, astro-ph/0101065 

\bibitem []{} Lightman, A.P., White, T.R., 1988, ApJ, 335, 57L

\bibitem[]{} Madej, J., R\'o\.za\'nska, A. 2000, A\&A 356, 654

\bibitem[]{} Matt, G., Fabian, A.C., Ross, R.R., 1993, MNRAS, 264, 839

\bibitem[]{} McKee C.F., Begelman M.C., 1990, ApJ, 358, 392

\bibitem[]{} Nayakshin, S., 2000, ApJ, 534, 718

\bibitem[]{} Nayakshin, S., Kazanas, D., Kallman, T.R., 2000, ApJ, 537, 833

\bibitem[]{} Nayakshin, S., Kallman, T.R., 2001, ApJ, 546, 406

\bibitem []{} Pequignot D., et al., 2001, in ``Spectroscopic 
Challenges of Photonized Plasma'', Eds. G. Ferland \& D. Savin, 
ASP Conference Series Vol. XXX


%\bibitem []{} Piro, L., Ba\l uci\' nska-Church, M., Fink, H., Fiore, F., 
%Matsuoka, M., Perola, G.C., Soffitta, P., 1997, A\&A, 319, 74 Sy1 
%E1615+061

\bibitem []{} Pojma\' nski, G., 1986, Acta Astr., 36, 69

\bibitem[]{}Poutanen J., 1999, in ``The Theory of
      Black Hole Accretion Discs'', eds. M.A. Abramowicz, G. Bj\'ornson,
      and J.E. Pringle, Cambridge University Press 

\bibitem{} Raymond, J.C., 1993, ApJ, 412, 267

\bibitem{} Reeves, J.N., Turner, M.J.L., Pounds, K.A.,
 O'Brien, P.T., Boller, Th., Ferrando, P.,
 Kendziorra, E., Vercellone, S., 2001, A\&A, 365, L134

\bibitem []{} Reynolds, C., 2000, in Probing the Physics of Acive Galactic 
Nuclei by Multiwavelength Monitiring, ASP Conference Series, vol. TBD, eds.
B.M. Peterson, R.S. Polidan, and R.W. Pogge (astro-ph/0009503)

\bibitem []{} Ross, R.R. 1978, PhDT.

\bibitem []{} Ross, R.R., Fabian, A.C., Mineshige, S., 1992, MNRAS, 258, 
189

\bibitem []{} Ross, R.R., Fabian, A.C., 1993, MNRAS, 261, 74 


\bibitem[]{} R\'o\.za\'nska, A. 1999, MNRAS 308, 751

\bibitem{} R\' o\. za\' nska, A., Czerny, B., 1996, Acta Astr., 46, 233  
[RC96] 

\bibitem{} R\' o\. za\' nska A., Czerny B., \. Zycki P.T., Pojma\' nski G.,
   1999, MNRAS, 305, 481

\bibitem{} Seaton, M.J., Yan, Y., Mihalas, D., Pradhan, A.K., 1994, MNRAS, 
266, 805

\bibitem[]{} Shakura N.I., Sunyaev R.A. 1973, A\&A, 24, 337


\bibitem []{} Shimura, T., Mineshige, S., Takahara, F., 1995, ApJ, 439, 74 

%\bibitem []{} Shimura, T., Takahara, F., 1993, ApJ, 419, 78 

\bibitem []{} Sincell, M.W., Krolik, J.H., 1997, ApJ, 476, 605

\bibitem []{} Sincell, M.W., Krolik, J.H., 1998, ApJ, 496, 737


\bibitem[]{} Suleimanov, V., Meyer, F., Meyer-Hofmeister, E., 1999, A\&A
350, 63

\bibitem[]{} Turner T.J., Romano P. George I.M., Edelson R., 
Collier S.J., Mathur, S., Peterson, B.M., 2001, ApJ, 561, 131 


\bibitem[]{} Yaqoob, T., George, I. M., Nandra, K.,
 Turner, T. J., Serlemitsos, P. J., Mushotzky, R. F., 2001, ApJ, 546, 759


\bibitem[]{} \. Zycki, P.T., Krolik, J.H., Zdziarski, A.A., Kallman, T.R., 
1994, 
   ApJ, 437, 597

\end{thebibliography}
\end{document}